\documentclass[useAMS,usenatbib]{mn2e}
\usepackage{graphicx,mathtools}
\usepackage{amssymb}
\usepackage[dvipsnames]{xcolor}
\usepackage{xparse,xcolor}
\usepackage{xspace}
\usepackage{cooltooltips}

\DeclareMathOperator{\dd}{{\rm d}}
\DeclareMathOperator{\id}{{\rm d}\!}
\newcommand{\der}[2]{\dfrac{\dd #1}{\dd #2}}
\newcommand{\pder}[2]{\dfrac{\partial #1}{\partial #2}}
\newcommand{\e}[1]{\!\times\! 10^{#1}}

\newcommand{\ah}{\mbox{$\alpha_{\mathrm{h}}$}\xspace}
\newcommand{\ac}{\mbox{$\alpha_{\mathrm{c}}$}\xspace}
\newcommand{\Scp}{\mbox{$\Sigma^{+}_{\mathrm{crit}}$}\xspace}
\newcommand{\Scm}{\mbox{$\Sigma^{-}_{\mathrm{crit}}$}\xspace}

\newcommand{\QSc}{\mbox{$Q\Sigma_{\mathrm{crit}}$}\xspace}
\newcommand{\Tcp}{\mbox{$T^{+}_{\mathrm{crit}}$}\xspace}
\newcommand{\Tcm}{\mbox{$T^{-}_{\mathrm{crit}}$}\xspace}
\newcommand{\Teffp}{\mbox{$T^{+}_{\mathrm{eff,crit}}$}\xspace}
\newcommand{\Teffm}{\mbox{$T^{-}_{\mathrm{eff,crit}}$}\xspace}
\newcommand{\Msun}{\mbox{$\mathrm{M_{\odot}}$}\xspace}

\newcommand{\Rout}{\mbox{$R_{\rm out}$}\xspace}
\newcommand{\Teff}{\mbox{$T_\mathrm{eff}$}\xspace}
\newcommand{\Tc}{\mbox{$T_\mathrm{c}$}\xspace}
\newcommand{\Mtr}{\mbox{$\dot{M}_{\mathrm{\rm tr}}$}\xspace}

\newcommand{\aML}{\mbox{$\alpha_{\mathrm{ml}}$}\xspace}
\newcommand{\Sig}{\mbox{$\Sigma$}\xspace}

\newcommand{\mean}[1]{\left< #1 \right>}
\def\bnabla{\mathbf{\nabla}}

\title[Dwarf Nova Outbursts with MRI Turbulence]
{Dwarf Nova Outbursts with Magnetorotational Turbulence}
\author[M. S. B. Coleman et al. ]{M. S. B. Coleman$^{1}$\thanks{E-mail:
mcoleman@physics.ucsb.edu}, I. Kotko${^2}$\thanks{Now at the Pennsylvania
State University, 222A Computer Building, University Park, PA 16802, USA},
O.  Blaes$^{1}$, J.-P.  Lasota$^{2,3}$, and S. Hirose$^{4}$\\
$^{1}$ Department of Physics, University of California, Santa Barbara, CA
93106, USA\\
$^{2}$ Nicolaus Copernicus Astronomical Center, Polish Academy of Sciences, Bartycka 18, 00-716 Warszawa,
Poland\\
$^{3}$ Institut d'Astrophysique de Paris, CNRS et Sorbonne Universit\'es, UPMC Univ Paris 06, UMR 7095, 98bis Bd Arago, 75014 Paris,
France\\
$^{4}$ Department of Mathematical Science and Advanced Technology, Japan
Agency for Marine-Earth Science and Technology, Yokohama, \\
\hspace{1cm} Kanagawa 236-0001, Japan
}

\begin{document}

\date{Accepted ---. Received ---; in original form ---}

\pagerange{\pageref{firstpage}--\pageref{lastpage}} \pubyear{2015}

\maketitle

\label{firstpage}

\begin{abstract}
The phenomenological Disc Instability Model has been successful in
reproducing the observed light curves of dwarf nova outbursts by invoking
an enhanced Shakura-Sunyaev $\alpha$ parameter $\sim0.1-0.2$ in outburst
compared to a low value $\sim0.01$ in quiescence.  Recent
thermodynamically consistent simulations of magnetorotational (MRI) turbulence
with appropriate opacities and equation of state for dwarf nova accretion
discs have found that thermal convection enhances $\alpha$ in discs in
outburst, but only near the hydrogen ionization transition.  At higher
temperatures, convection no longer exists and $\alpha$ returns to the
low value comparable to that in quiescence.  In order to check whether
this enhancement near the hydrogen ionization transition is sufficient to
reproduce observed light curves, we incorporate this MRI-based variation in
$\alpha$ into the Disc Instability Model, as well as simulation-based
models of turbulent dissipation and convective transport.
These MRI-based models can successfully reproduce observed outburst and
quiescence durations, as well as outburst amplitudes, albeit with different
parameters from the standard Disc Instability Models.  The MRI-based model
lightcurves exhibit reflares in the decay from outburst, which
are not generally observed in dwarf novae. However, we highlight the problematic
aspects of the quiescence physics in the Disc Instability Model and MRI simulations that are responsible for this
behavior.
\end{abstract}


\section{Introduction}
\label{sec_intro}

Dwarf novae are transient optical outbursts observed from close binary systems
containing an accreting white dwarf.  These outbursts
have amplitudes up to $\sim 8\,\,\mathrm{mag}$ and last 2-20 days, with
recurrence times ranging from $\sim 4$ days to years \citep{L01}.
The Disc Instability Model (hereafter DIM) is a well-tested model that
attributes the outbursts to a thermal-viscous\footnote{To be clear,
``viscosity" here and throughout the paper is an effective turbulent
viscosity as opposed to a true molecular viscosity.}
instability that arises at temperatures where hydrogen is ionizing in the
accretion disc.
During quiescence, a cold dwarf nova disc accumulates matter and heats until
somewhere the temperature crosses a critical value which triggers the thermal
instability. This creates heating fronts which propagate into the
low-temperature zones, leaving behind ionized regions.
In these hot regions of the disc, the turbulence and the \citet{SS73}
$\alpha$ parameter are assumed to be enhanced, leading to an increase in angular
momentum transport.  This causes mass, which during quiescence had gathered mostly in
the outer parts of the disc, to diffuse inwards at a high rate.
The DIM is also successful in explaining soft X-ray transient outbursts observed
in close binary systems containing accreting neutron stars and black holes,
although X-ray irradiation of the outer disc modifies the stability criterion
\citep{VP96} and plays an important role in
extending the duration of the outburst in these systems \citep{KIN98,DUB01}.

In addition to their intrinsic interest,
the observed amplitudes and time scales present in dwarf nova light
curves provide the best quantitative measurements of the stresses responsible
for angular momentum transport in accretion discs \citep{KIN07}.
It was realized early on (e.g. \citealt{MO,SMAK84, MMH84}) that the observed
outburst amplitudes can only be reproduced in the DIM if the stress parameter
$\alpha$ takes on different values in the outburst (denoted by h for hot)
and quiescent (denoted by c for cold) states, with \ah being larger than \ac
by about a factor of ten.  Modern versions of the DIM use an interpolation
scheme between constant values of \ah and \ac in the outburst and
quiescent states, but this does not imply that $\alpha$ is necessarily
constant for all high or for all low temperatures\footnote{E.g.  \citet{MW} used
$\alpha\sim \left(H/R\right)^{1.5}$.}.  Detailed fits of the DIM
to the observed correlations between orbital period (or, equivalently, outer
disc radius) and the outburst
duration and decay rate
in normal, U~Gem-type
systems and normal outbursts of SU UMa systems, give values for \ah of between
0.1 and 0.2, and certainly rule out
significantly smaller values \citep{SMAK99,K12}.
Moreover, only such high
values are able to explain the observed
linear relationship between
outburst amplitude and the logarithm of outburst recurrence time
(the Kukarkin-Parenago relation; \citealt{K12}).

Ever since the first application of the magnetorotational instability (MRI)
to accretion discs \citep{BAL91,HAW91}, it has been widely suspected that
both angular momentum
transport and dissipation of mechanical energy is mediated by MRI turbulence,
at least if the disc is sufficiently electrically conducting.  The accretion
stress is due to correlations between radial and azimuthal magnetic field
fluctuations, as well as radial and azimuthal velocity fluctuations
(e.g. \citealt{BAL98}).  One can measure this stress directly from numerical
simulations of the turbulence.
Equivalent values of the $\alpha$ parameter can then be derived by scaling
this stress with the average thermal pressure.

Because the values of $\alpha$ are among the most fundamental ways of
confronting MRI turbulence with observations, many groups have measured
these values from their simulations.  In the absence of net vertical magnetic
flux, local shearing box simulations that incorporate vertical gravity
generally give time-averaged $\alpha$ values of around two or three percent
\citep{HIR06,DAV10,SHI10,GG11,SIM12}.  Global simulations without net vertical
flux of the entire disc,
but which allow for large scale field loops that can produce local
vertical magnetic fluxes threading the disc, also so far produce
comparably small values of $\alpha$ \citep{SOR10,HAW11,SOR12},
though more
work needs to be done in exploring the effects of various field topologies.

A number of suggestions have been made as to how to resolve the discrepancy
between the high values of $\alpha$ inferred from dwarf novae
in outburst,
and these low values measured in MRI simulations.
First, it has been known for some time that shearing box simulations with
no vertical gravity have stronger $\alpha$ values when the box is threaded
by a vertical magnetic field, and the resulting $\alpha$ increases
with magnetic field strength so long as the critical vertical wavelength
of the MRI lies within the box \citep{HAW95,SAN04,PES07}.\footnote{Recent
work by \citet{SHI16} has also shown that the value of $\alpha$ in
shearing boxes with no vertical magnetic field or vertical gravity is
sensitive to the height of the box:  taller boxes produce significantly
higher values of $\alpha$.}  Because the
initial total vertical magnetic flux is necessarily conserved in shearing
box simulations, this would appear to require imposing a net external
vertical magnetic field in the disc from the outside in order to increase
the value of $\alpha$ in outburst states. This might result from the magnetic
field of the companion star, although one would then have to explain why
the resulting value of $\alpha$ is so universal in the outburst state.

On the other hand, the increase
in local stress with {\it local} vertical magnetic flux has been confirmed in
global simulations with no overall net vertical magnetic flux, although
the global disc still has a low value of $\alpha$ when averaged over the entire
disc \citep{SOR10}.  Having net
vertical flux may also drive magnetocentrifugal winds, which
can also extract angular momentum from the disc \citep{SUZ09,FRO13,LES13,BAI13}.
In addition to the vertical magnetic field, transient phases such as those caused
by the dwarf nova
outbursts themselves may also produce periods of enhanced $\alpha$
that are similar to that observed in transient magnetic field growth phases
in global simulations \citep{SOR12}.

Even without vertical magnetic field or transient behavior, however, radiation
MHD stratified shearing box simulations that incorporate a realistic equation
of state and opacities near the hydrogen ionization transition reproduce
thermal equilibria resembling the stable upper and lower branches of
the local ``S-curve" of the DIM \citep{HIR14}.  As in previous MRI simulations,
$\alpha$ values of a few percent generally result, but $\alpha$ dramatically
increases as one approaches the lower end of the upper branch.  This appears to
be due to the onset of intermittent vertical convection caused by the large
increase in opacity near the ionization transition.  The resulting vertical
motions at the beginning of the convective episodes build up vertical
magnetic field, which may be seeding the axisymmetric MRI. In addition,
temporal phase differences between the variations of stress and pressure may also be
increasing the time-averaged $\alpha$.\footnote{Persistent convection can
also enhance $\alpha$, provided the Mach numbers of the convective motions
exceed $\sim0.01$ \citep{HIR15}.}  These simulations therefore
reproduce the observed high values of $\alpha$ on the upper branch
of the S-curve, but only near the low-temperature end of the upper branch.

In addition to the variations of $\alpha$ on the upper branch, there are
significant differences in the time-averaged vertical structure observed
in the simulations and the standard assumptions used in the DIM.  First,
the DIM generally assumes that the stress to pressure ratio is
constant with height, so that the dissipation rate per unit volume is
proportional to pressure.  MRI simulations, on the other hand, generally
produce a more extended vertical dissipation profile \citep{TUR04,HIR06}.
Moreover, the simulated upper, near-photosphere
layers are generally supported against the vertical tidal gravity by magnetic
forces, not thermal pressure forces \citep{MIL00,HIR06}, in contrast to
the DIM vertical structure models.
Finally, as we have just discussed,
vertical transport of heat in the simulations is caused by alternating
episodes of radiative diffusion and thermal convection, at least near the
end of the upper branch. The DIM incorporates the possibility of convection
using mixing length theory, but it is not clear that this prescription
adequately describes what is happening in the MRI simulations.

The primary purpose of this paper is to incorporate the variation of
$\alpha$ measured in the MRI simulations into the DIM, to see whether an
enhancement of $\alpha$ just near the end of the upper branch is sufficient
to reproduce the observed amplitudes and time scales in dwarf nova outburst
light curves.  A secondary objective is to use the dissipation
and heat transport observed in the simulations to produce more physics-based
vertical structure models that can be incorporated in the DIM.
In section 2, we discuss the behavior of the $\alpha$ parameter in the MRI
simulations, including new simulations done at different radii in the
disc from those originally done by \citet{HIR14}.  It turns out that $\alpha$
can be reasonably fit by a function of just local disc effective temperature.
In section 3 we present time-averaged vertical dissipation profiles
from the simulations, and show how these can be incorporated into the DIM
vertical model equations.  We also briefly discuss the evolution of MRI simulations
when they are not in thermal equilibrium, and show how this evolution is
surprisingly consistent with the standard effective $\alpha$ prescription
currently used in the DIM.  Finally, we also present a mixing length
prescription which appears able to reproduce the time-averaged vertical
structures observed in the simulations on the upper branch.  In section 4,
we present theoretical
dwarf nova outburst light curves using these MRI simulation-based
prescriptions, and
compare them to the standard DIM light curves.  We discuss our results
in section 5, and summarize our conclusions in section 6.

\section{MRI Simulation-Based Stress Prescription}
\label{sec_simulations}

As discussed in the introduction, the ratio of vertically averaged stress
to vertically averaged thermal pressure, $\alpha$, can be inferred from
observations and can be measured directly from simulations.
The difference in $\alpha$ between the lower and upper branches of the S-curve
has been crucial for getting DIMs to produce realistic light curves. In the
standard version of the
DIM two values of alpha were used;  \ah$\sim 0.1$ for the hot, ionized
upper branch, and \ac, 4 to 10 times smaller, for the lower branch, with
some smooth and rapid transition between them.  In particular, our modeling
of the standard DIM here in this paper will use a slightly modified version
of the prescription first introduced by \citet{H98}:
\begin{equation}
\label{eqn:alphadim}
\log(\alpha)=\log(\ac)+[\log(\ah)-\log(\ac)]\left[1+\left(
\frac{T_{\mathrm{c0}}}{\Tc}\right)^{16}\right]^{-1},
\end{equation}
where \Tc is the central (midplane) temperature and
$T_{\mathrm{c0}}$ is a critical temperature.  The choice of this parameter
is somewhat arbitrary and we take $T_{\mathrm{c0}}=2.9\times10^4$~K
\citep{LDK}, and take the limiting $\alpha$-parameters to be
$\alpha_{\rm c}=0.03$ and $\alpha_{\rm h}=0.12$.

Until recently,
local shearing box MRI simulations (without net vertical flux through
the simulation domain)
produced values
of $\alpha$ which are too low for the hot ionized outburst state of dwarf
novae. However, for the first time, $\alpha$
measured in local MRI simulations \citep{HIR14} appear to be consistent
with $\alpha$ values inferred from observations and commonly used to reproduce dwarf novae outbursts
by the DIM.  \citet{HIR14} found
that the $\alpha$ parameter varies along the S-curve and is not constant on the
upper branch, with an enhancement of $\alpha$ towards the tip (low temperature
end) of the upper branch.
This provides an intriguing replacement for the previous ad hoc bimodal
$\alpha$ prescription previously used \citep[e.g.][]{H98, L01, K12}.

The \citet{HIR14} simulations are done in the geometry of a vertically
stratified shearing box, in which a small local patch of an accretion disc is
approximated
as a co-rotating Cartesian frame $(x,y,z)$ with linearized Keplerian shear flow
$-(3/2)\Omega x$, where $x$, $y$, and $z$ correspond to the
radial, azimuthal, and vertical coordinates, respectively \citep{HAW95}.
The simulations assume no explicit shear viscosity in the basic equations,
and no Ohmic, Hall, or ambipolar diffusion effects are included either, i.e.
ideal MHD is assumed.  The simulations are therefore probably not accurate
for the quiescent, largely electrically neutral state
where Ohmic dissipation and the Hall effect are likely important (see Appendix \ref{sec:non-ideal}).
Magnetic and kinetic
energy losses at the grid scale are captured and added to the local internal
energy of the gas, creating an effective turbulent dissipation
\citep{T03,HIR06}.  Additionally, the simulations include realistic equation of
state and opacity tables in order to accurately model
the hydrogen ionization regime \citep[see][for details]{HIR14}, and
the flux limited diffusion approximation
(which breaks down at low optical depths) is used to model radiation transport and cooling.

The simulations in \citet{HIR14} were computed for only one angular frequency
$\Omega = 6.4\times10^{-3}$ s$^{-1}$, which corresponds to a
distance\footnote{Due to some rounding differences our distance slightly differs
from the $1.23\e{10}$~cm that \citet{HIR14} state.} of $1.25\e{10}$~cm from a
white dwarf of $0.6 M_\odot$. In order to explore a larger parameter space, we
utilize the same methods to run simulations at two additional orbital radii,
$1.25\e{9}$~cm and $4.13\e{9}$~cm. The parameters for these additional simulations
can be found in Table \ref{table:sims}.

We incorporate simulation data into the DIM's vertical thermal equilibrium
equations \citep{H98} by first discarding the initial 10 orbits of data (this is the time it take for turbulence to develop and erase details of the initial conditions) and taking a time average of the remaining duration of the simulations. All the simulations have been run long enough so that at least 100 orbits (and up to 220 orbits) of data are included in the time averaging. This ensures that averaging is done over many
thermal times so that thermal fluctuations are smoothed out and that the average is well behaved (i.e. running the simulation for another 10 orbits would have little effect on the time average).
We also
horizontally average the data over the radial ($x$) and
azimuthal ($y$) directions.  Finally, we vertically symmetrize the data about
the midplane $z=0$. Simulation data which have
undergone these operations will henceforth be referred to as profiles. To
summarize, for some scalar data $f$ the profile is calculated as follows
\begin{align}
f(z)\equiv\dfrac{\int\!\!\int\!\!\int \left[f(x,y,z,t)+ f(x,y,-z,t)\right] \,\id x\id y\id t}{2\int\!\!\int\!\!\int \id x \id y \id t}.
\end{align}
We then fit some of these profiles in order to produce vertical structures
that more accurately represent simulation physics than the standard DIM.

The parameter $\alpha$ is computed from each simulation as follows
\begin{equation}
\alpha\equiv \dfrac{\int w_{xy}(z) \id z}{\int \left[P_{\rm gas}(z) + P_{\rm rad}(z)\right] \id z},
\end{equation}
where $w_{xy}$ is the sum of the Maxwell and Reynolds stresses
\begin{equation}
w_{xy}\equiv -\dfrac{B_xB_y}{4\pi}+\rho v_x\delta v_y.
\end{equation}
Here $B_x$ and $B_y$ are the radial and azimuthal components of the
magnetic field respectively\footnote
{Note that the magnetic fields as defined here differ by a factor of
$\sqrt{4\pi}$ from the magnetic fields defined in the simulations.  Here
we use standard cgs Gaussian units.},
$v_x$ is the radial velocity, and $\delta v_y\equiv v_y+3\Omega x/2$
is the difference between the azimuthal
velocity and the mean rotational velocity in the disc.
We also examined how the temporal mean of $\alpha$ varies with time and we found these fluctuations diminished with time,
indicating that the time averaged $\alpha$ is well behaved.  Provided the
time average is done over an interval of at least 100 orbits, differences
in the time-averaged value of $\alpha$ within a single simulation are
less than ten percent.

\begin{figure}
\centering
\includegraphics[width=\linewidth]{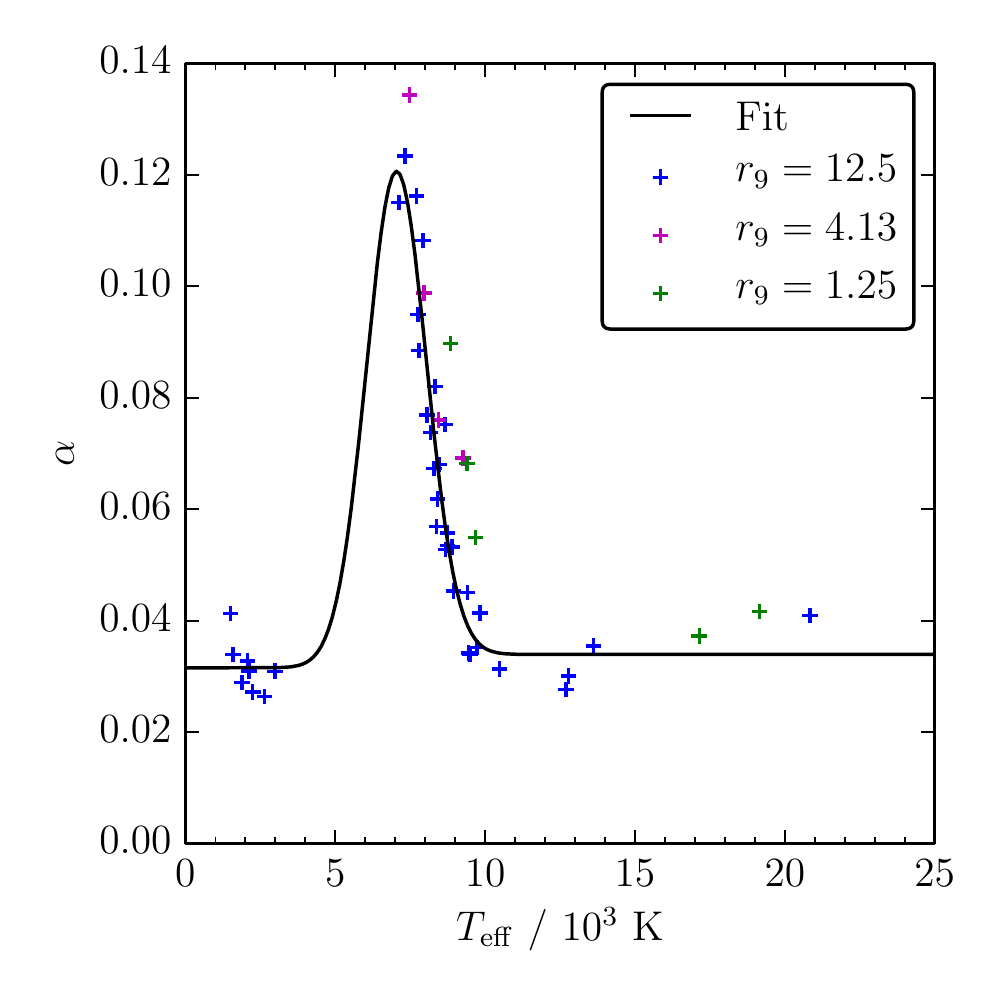}
\caption{Time averaged $\alpha$ versus effective temperature for all the MRI simulations
at three different radii around a $0.6M_\odot$ white dwarf: $r=1.25\e{10}$ cm (blue crosses),
$r=4.13\e{9}$ cm (magenta crosses), and $r=1.25\e{9}$ cm (green crosses). The best fit curve
for the $r=1.25\e{10}$ cm simulations is plotted in black.
}
\label{fig:aT}
\end{figure}

We fit the variation of $\alpha$ between all the simulations at radius
$r=1.25\times10^{10}$~cm  with a prescription depending only on local effective
temperature $T_{\rm eff}$ of the following form
\begin{align}
\label{eqn:alphamri1}
\alpha &= a\exp\left(-\dfrac{x^2}{2}\right)+\dfrac{b}{2}\tanh\left(x\right)+c,\\
\label{eqn:alphamri2}
x&=\dfrac{T_{\rm eff}-T_0}{\sigma}.
\end{align}
The best fit parameter values are $a=8.79\e{-2}$, $b=2.41\e{-3}$,
$c=3.27\e{-2}$, $T_0=7034$ K, and $\sigma=1000$ K.  Figure \ref{fig:aT}
compares this fit to the simulation data at this radius.  Also shown are
the results from simulations at two additional radii (which are not included in the fit), which are also
reasonably consistent with this fit, though there is some indication that
the simulations at smaller radii have slightly greater convective enhancements
of $\alpha$.  We will use the fit of equations (\ref{eqn:alphamri1}) and
(\ref{eqn:alphamri2}) in the simulation-based light curve modeling below.

\section{MRI Simulation-Based Vertical Structure Models}

In addition to the overall behavior of the stress to pressure ratio $\alpha$
discussed in the previous section, MRI simulations also exhibit differences
with the standard DIM assumptions concerning the local vertical structure of
the disc.  Here we utilize data from the \citet{HIR14} vertically stratified
shearing box MHD simulations to show how to make DIM vertical structure models
that better reflect some of the actual properties of the turbulence observed in
the simulations.

\subsection{Dissipation Profile}

Previously the DIM has relied on an ad hoc assumed vertical profile of turbulent
dissipation, in particular a {\it vertically local} $\alpha$ prescription in
which the dissipation rate per unit volume is proportional to the local
thermal pressure.  This is
very different
from what is observed in vertically
stratified MRI simulations.  Such simulations of course cannot spatially
resolve the actual microscopic viscous and resistive length scales in the
plasma, and the simulations of \citet{HIR14} do not include any explicit
resistivity or shear viscosity in the basic equations.  Instead, a total
energy scheme is employed in which grid scale losses of magnetic and
kinetic energy are automatically transferred to internal energy of the
plasma, thereby effecting a dissipation of turbulent mechanical energy.
This will capture the true dissipation rate in the turbulence provided
the turbulent cascade that would exist in reality below the grid scale
is capable of transferring most of the energy down to the true microscopic
dissipation scales.  Whether this accurately captures the dissipation
occurring in real discs in nature remains to be seen.

In fact, there is strong numerical evidence that the saturation
level of the turbulent stresses, and even whether long-lived turbulence
can be maintained, depends on the values of the fluid and magnetic Reynolds
numbers or their ratio, the magnetic Prandtl number.  This is particularly
true of shearing box simulations that lack vertical gravity (e.g.
\citealt{FRO07, LES07}).  Including vertical gravity appears to slightly
extend the range of magnetic Prandtl numbers that allow sustained turbulence
to lower (but still greather than unity) values \citep{DAV10}.  This is
also true of shearing box simulations which lack vertical gravity, provided
the box height is large enough \citep{SHI16}.
Because the simulations
used here have no explicit viscosity or resistivity, and dissipation
effectively occurs at the grid scale, the effective magnetic Prandtl number
must be of order unity.  While the simulations nevertheless exhibit sustained
turbulence, more work needs to be done to investigate whether and how the
dissipation profiles, and even the saturation level of the turbulence, might
be affected by the actual dissipation scales.  Dependencies of the
turbulent stresses on magnetic Prandtl number may even themselves lead
to thermal instabilities in accretion discs \citep{POT14}.

\begin{figure}
\begin{center}
\includegraphics[width=\linewidth]{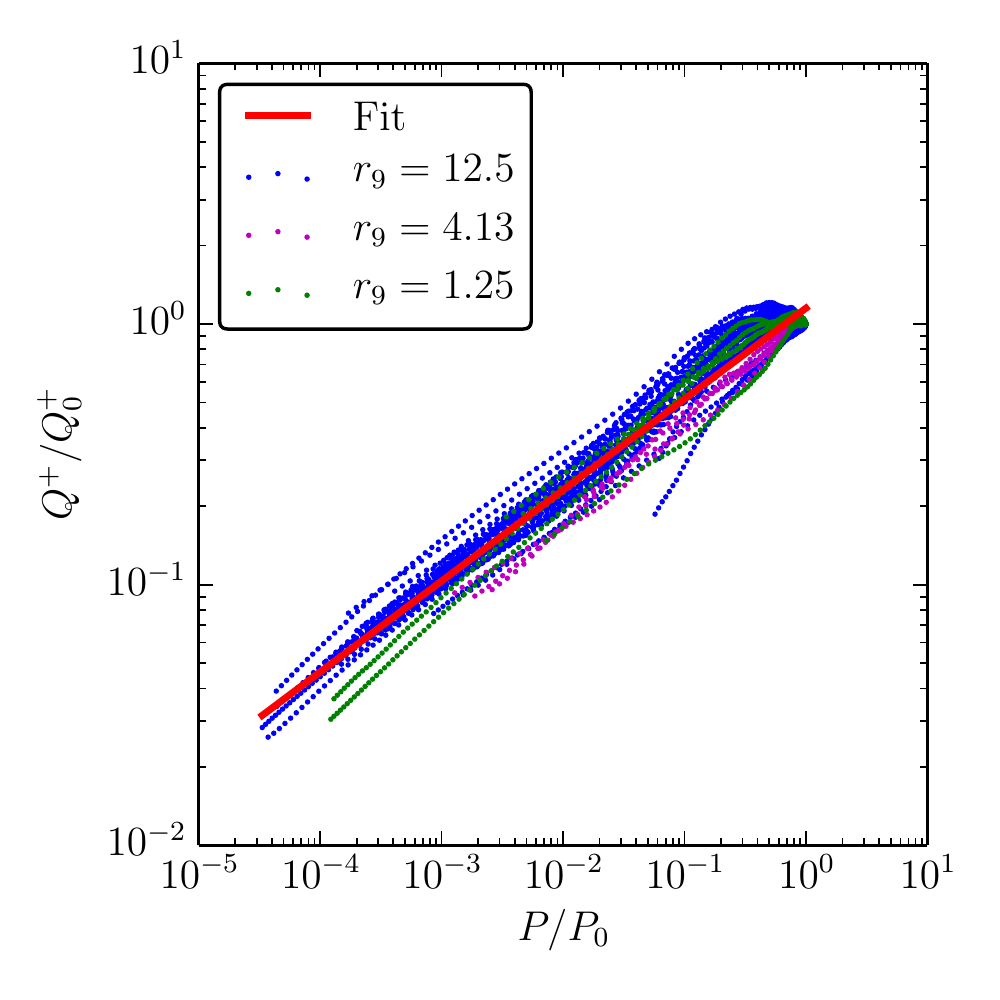}
\end{center}
\caption{
Time and horizontally averaged profiles of dissipation rate per unit volume
as a function of thermal pressure, each scaled by their respective midplane
values, for each of the MRI simulations used in this paper.  The different colors refer to
simulations done at three different radii around a $0.6 M_\odot$ white dwarf:
$r=1.25\e{10}$ cm (blue points), $r=4.13\e{9}$ cm (magenta points), and $r=1.25\e{9}$ cm
(green points).  A power law fit to all the profiles is also plotted (red line). Note
that the slope of the fit is significantly more important than the vertical offset,
which we determine in our light curve modeling by enforcing energy conservation
using equation (\ref{eqn:SSalpha}).}
\label{fig:Qfit}
\end{figure}

In any case, as shown in Figure~\ref{fig:Qfit}, the time and
horizontally-averaged vertical profile of dissipation rate per unit volume
in the simulations
is not proportional to
the pressure, in contrast to the simple
assumption used in the DIM. In fact, the vertical dissipation profile
generally does not even decline monotonically away from the midplane, but
instead peaks off the midplane, possibly due to the effects of magnetic
buoyancy \citep{BLA11}.
Despite this non-monotonic behavior, we find
that we can adequately replace the usual DIM assumption of dissipation
rate per unit volume $Q^+$ being simply proportional to thermal pressure
with, instead, a power law dependence on thermal pressure
(see Figure \ref{fig:Qfit}):
\begin{equation}
\label{eqn:Qfit}
\dfrac{Q^+}{Q_0^+}=\left(\dfrac{P}{P_{0}}\right)^\delta,
\end{equation}
where the subscript zero is used to denote midplane values.
By thermal pressure $P$, we mean the sum of gas pressure ($P_{\rm gas}$)
and the much smaller radiation pressure ($P_{\rm rad} \ll P_{\rm gas}$),
but we exclude magnetic pressure, which typically dominates the thermal
pressure far from the midplane.
Using linear regression (in log space) we determined that the best fit
exponent $\delta=0.35$.
The ratio $Q_0^+/P_0^{\,\delta}$
is chosen such that
the vertically integrated \citet{SS73} $\alpha$ relation holds:
\begin{equation}
\label{eqn:SSalpha}
\dfrac{3}{2} \alpha \Omega \int_{-\infty}^{\infty} P(z)\,\id z=
\int_{-\infty}^{\infty} Q^+(z)\,\id z.
\end{equation}

\subsection{Non-Equilibrium Dissipation}

During the passage of a heating or cooling front through the disc, annuli
at the location of the fronts are of course out of thermal equilibrium.
Within the DIM \citep{H98}, this is handled by constructing hydrostatic
vertical structures with a local vertical energy flux divergence given by
\begin{equation}
\label{eqn:alphaeffdim}
\frac{dF_z}{dz}=\frac{3}{2}\alpha_{\rm eff}\Omega P(z),
\end{equation}
where $P(z)$ is the local thermal pressure and $\alpha_{\rm eff}$ is a
parameter that differs from $\alpha$ when the annulus is not in thermal
equilibrium, both because vertically integrated heating and cooling will
then no longer be equal and because of the concomitant vertical thermal
expansion or contraction.  The actual value that $\alpha_{\rm eff}$
takes is determined by solving for the complete vertical structure
for a given effective temperature and surface mass density.
If thermal equilibrium does hold, then
$\alpha_{\rm eff}=\alpha$, and we recover the standard DIM assumption
that the dissipation rate per unit volume is proportional to the local
thermal pressure at every height in the annulus.

As just noted, the time and horizontally averaged vertical profiles of
dissipation rate per unit volume measured in the MRI simulations do not
simply scale with local thermal pressure. We therefore modify equation
(\ref{eqn:alphaeffdim}) to be consistent with our fit to the equilibrium
dissipation profile, equation~(\ref{eqn:Qfit}):
\begin{equation}
\label{eqn:alphaeffdimri}
\frac{dF_z}{dz}=\frac{3}{2}\alpha_{\rm eff}\Omega P_{0}
\left(\dfrac{P(z)}{P_{0}}\right)^\delta.
\end{equation}
However, it is not obvious that the MRI simulations should behave according
to this equation outside thermal equilibrium.

\begin{figure}
\begin{center}
\includegraphics[width=\linewidth]{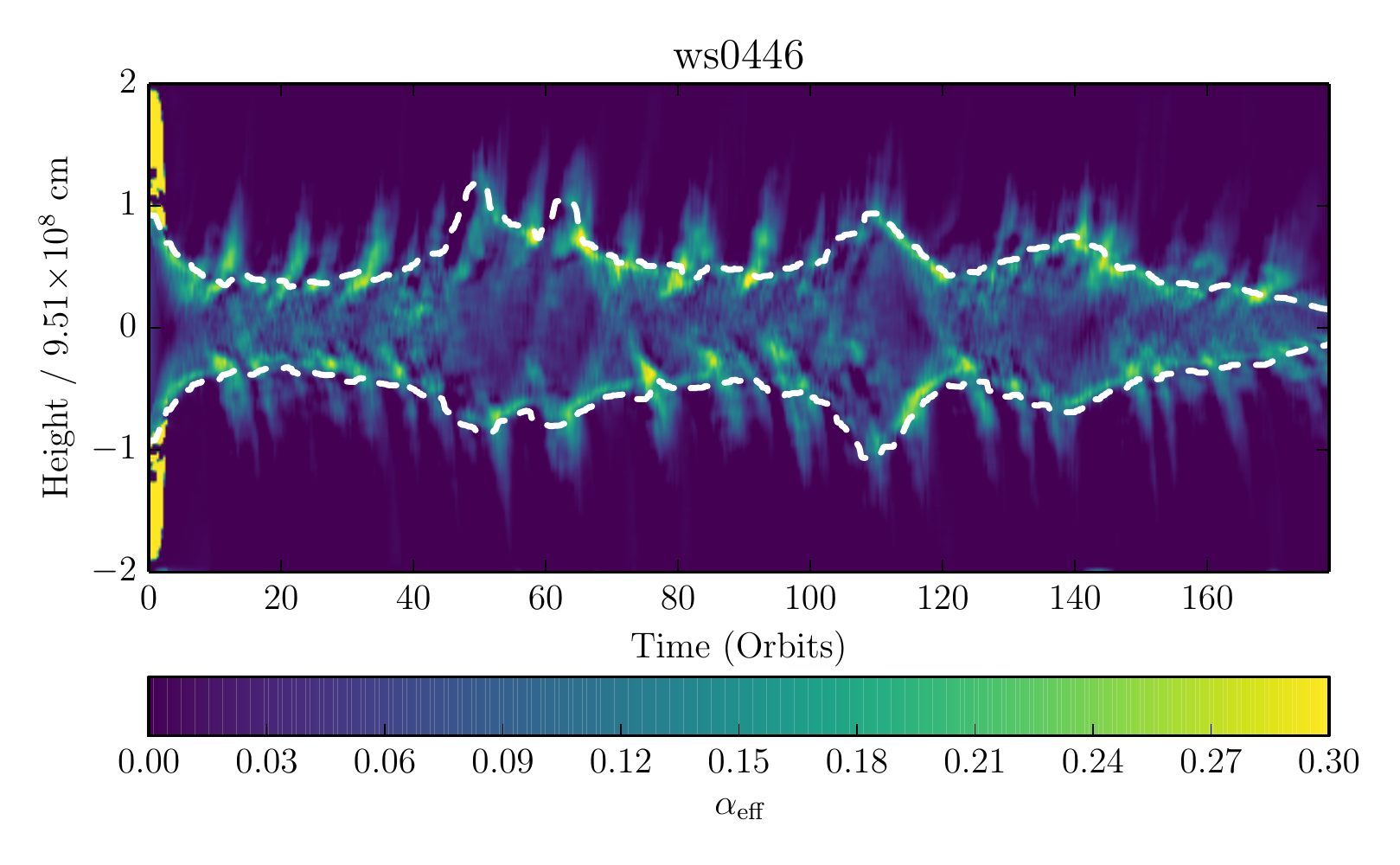}
\end{center}
\caption{Horizontally averaged $\alpha_{\rm eff}$ for simulation ws0446,
a stable convective simulation.  The horizontal axis is time in orbits, and
the vertical axis is height. The white dashed contours are the photospheres.
The data has been smoothed with a two orbit running boxcar.}
\label{fig:ws0446}
\end{figure}

\begin{figure}
\begin{center}
\includegraphics[width=\linewidth]{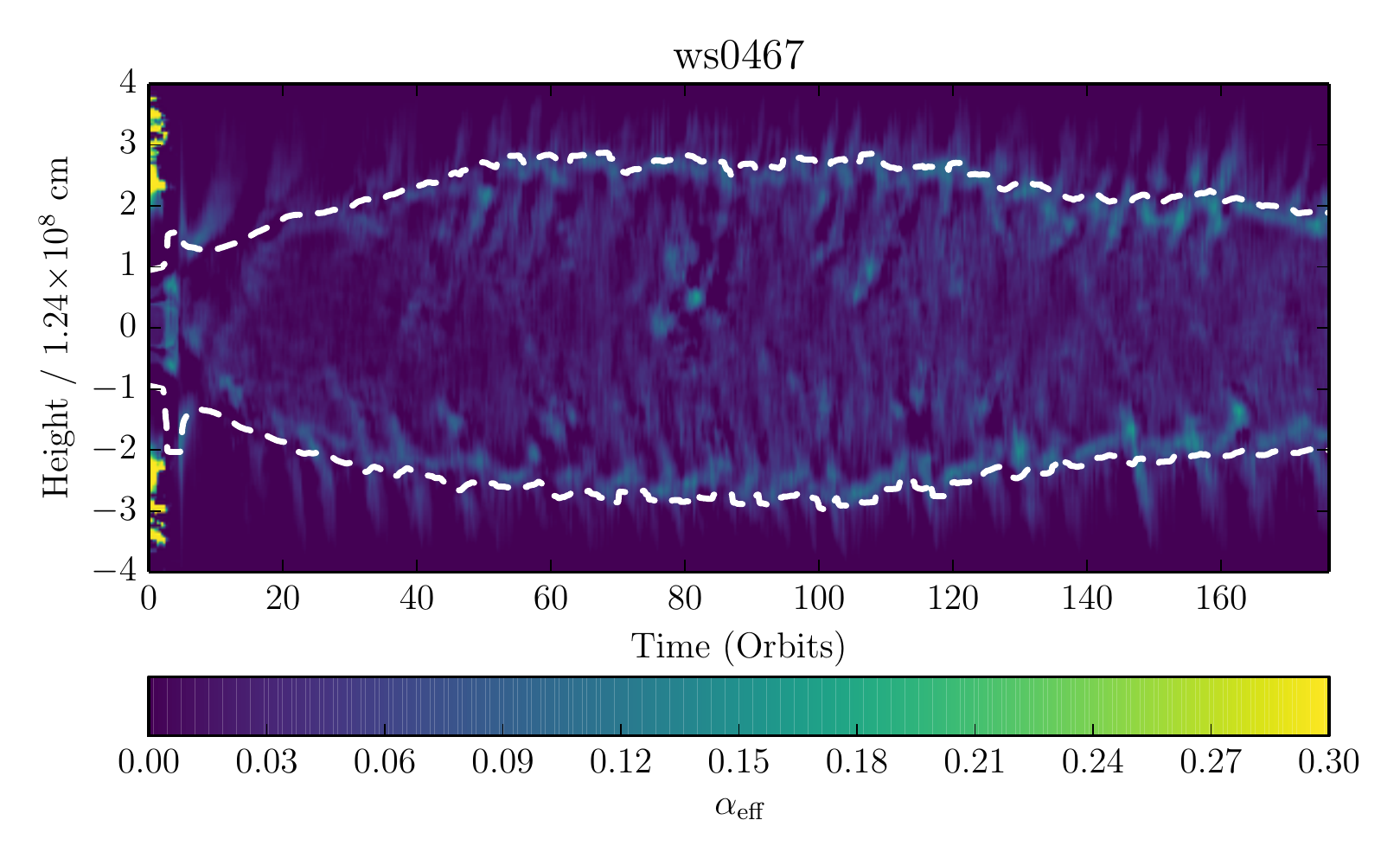}
\end{center}
\caption{Horizontally averaged $\alpha_{\rm eff}$ for simulation ws0467, an
unstable simulation that undergoes runaway heating until significant mass
loss occurs through the vertical boundaries. The horizontal axis is time in
orbits, and the vertical axis is height. The white dashed contours are the
photospheres. The data has also undergone two orbit boxcar smoothing.  This
simulation exhibits continuous convective vertical transport of heat.}
\label{fig:ws0467}
\end{figure}

\begin{figure}
\begin{center}
\includegraphics[width=\linewidth]{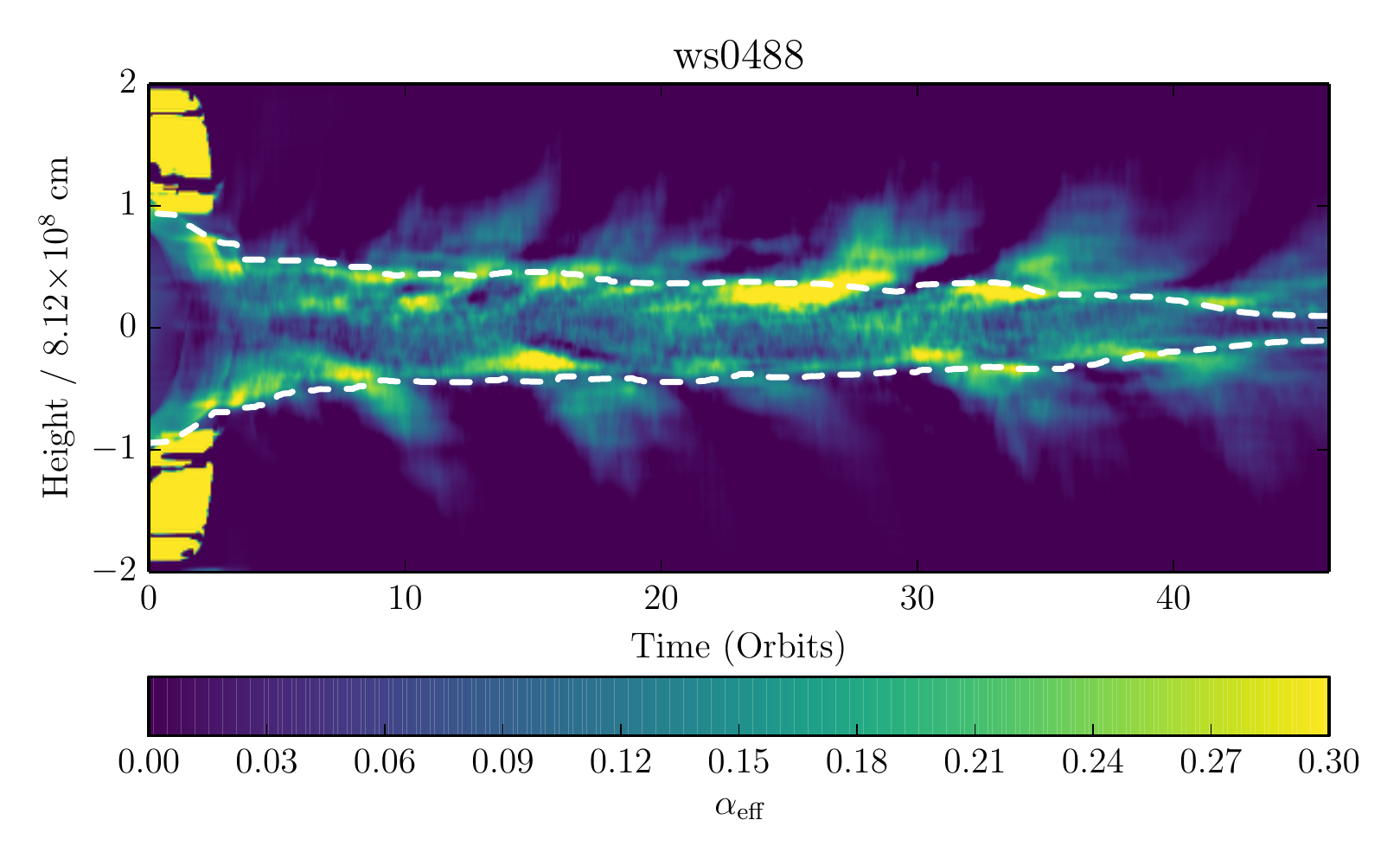}
\end{center}
\caption{Horizontally averaged $\alpha_{\rm eff}$ for simulation ws0488,
an unstable simulation that undergoes runaway cooling. The horizontal axis is
time in orbits, and the vertical axis is height. The white dashed contours are
the photospheres. The data has also undergone two orbit boxcar smoothing.}
\label{fig:ws0488}
\end{figure}

If the assumption of equation (\ref{eqn:alphaeffdimri}) were perfect, then all
simulation data would fall directly on the best-fit line in
Figure~\ref{fig:Qfit}. While time averaged profiles lie near this line, it is
not apparent that one should expect this behavior from a thermally evolving
simulation. To test this we defined
\begin{equation}
\alpha_{\rm eff}(z,t)\equiv \frac{2}{3}\frac{1}{\Omega P_{0}}
\frac{\partial F_z}{\partial z}\left(\frac{P_{0}}{P} \right)^\delta,
\end{equation}
with $\delta=0.35$ as previously discussed, and examined variations in
$\alpha_{\rm eff}$ for a few simulations
(see Figures \ref{fig:ws0446}-\ref{fig:ws0488}).
In addition to stable simulations (e.g. ws0446
shown in Figure \ref{fig:ws0446}), we specifically examined two
non-equilibrium simulations: one heating (ws0467, see Figure \ref{fig:ws0467}),
and one cooling (ws0488, see Figure \ref{fig:ws0488}). These simulations
were started just beyond the edges of the lower and upper branches of the
S-curve, respectively (see Figure 11 of \citealt{HIR14}).
As these two simulations
evolve, they move around on the plane of $T_{\rm eff}$ vs. total column
mass density $\Sigma$, which allows
$\alpha_{\rm eff}$ to vary with time.  However, $\alpha_{\rm eff}$ should be
approximately constant in height at a given time if equation
(\ref{eqn:alphaeffdimri}) is to adequately describe the simulation behavior.
With the exception of variations near and outside the photospheres, this seems
to be a reasonable approximation for simulations ws0446 and ws0467.
There are some clear issues for the cooling simulation ws0488 just below the photosphere (see Figure \ref{fig:ws0488}),
which manifest as asymmetric regions of enhanced $\alpha_{\rm eff}$.
It is possible that these regions arise from asymmetric cooling/collapsing of the disc,
which is not possible to incorporate into the DIM and show a clear limitation of our strategy.
However, outside these regions, ws0446 and ws0488 show comparable variations,
signifying that this approach is not unreasonable.

\subsection{Mixing Length Theory}

\begin{figure}
\centering
\includegraphics[width=\linewidth]{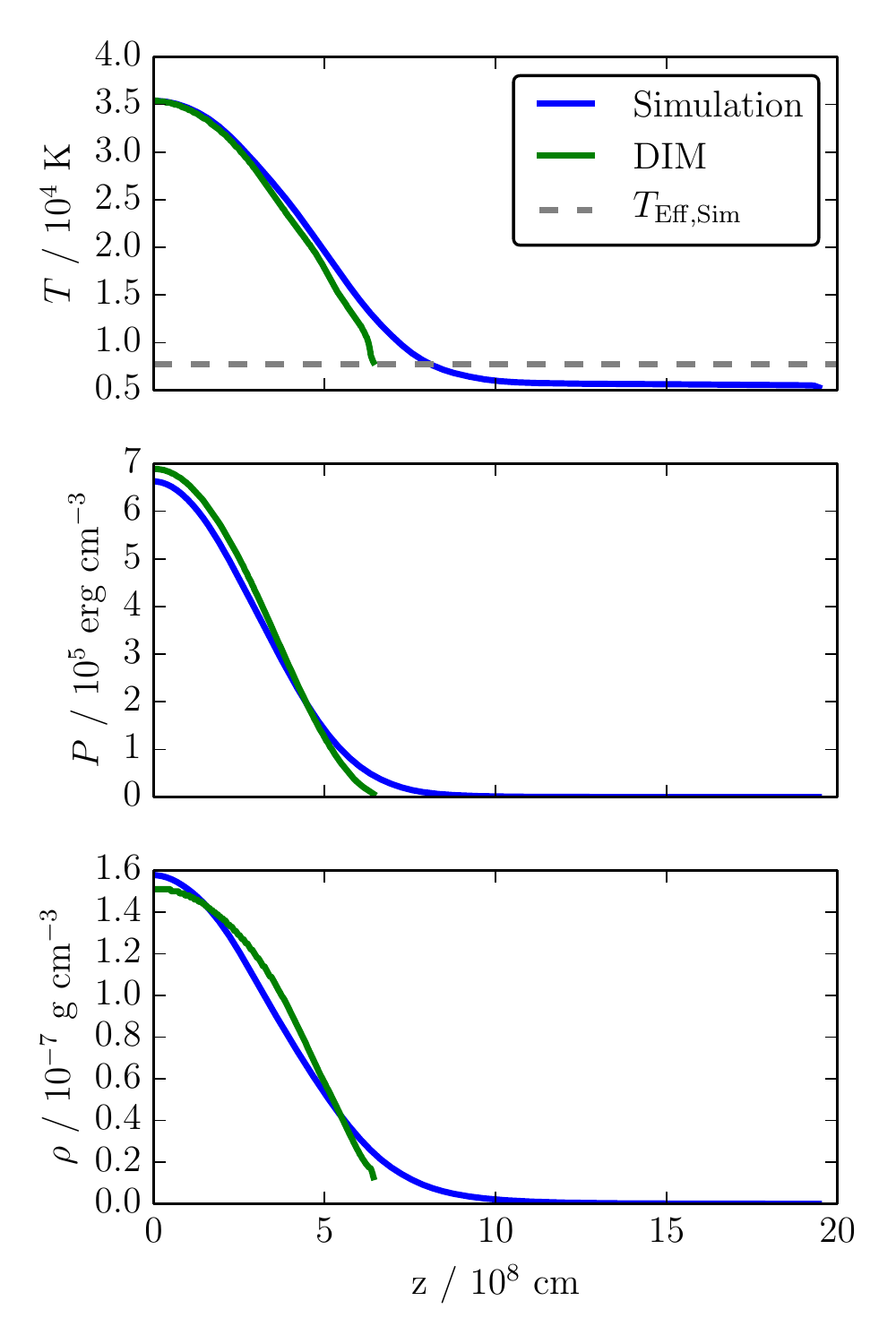}
\caption{
Comparison of vertical structure model to a convective simulation (ws0441) with $r=1.25\e{10}$ cm. Profiles of temperature (top), thermal pressure (middle), and mass density (bottom) are plotted verses $z$ for  profiles measured from the simulation (blue) and vertical structure models (green).
In these plots, it is clear that the simulations extend further than our models.
This is because our DIM vertical structures use the photosphere as a vertical boundary condition and thus terminate there. Thus the minimum temperature in the vertical structure model is also the effective temperature.
The time averaged effective temperature of the simulation is plotted as the dashed gray horizontal line.
Thus this figure clearly shows good agreement between our modified DIM and the simulations.
}
\label{fig:vert-mod}
\end{figure}

The vertical temperature profile is computed in the DIM according to
\begin{align}
\label{eqn:T-grad}
\nabla &\equiv \der{\ln T}{\ln P} =
\begin{cases}
\nabla_{\rm rad} & \text{if }\nabla_{\rm rad}\leq\nabla_{\rm ad}\\
\nabla_{\rm conv} & \text{if }\nabla_{\rm rad}>\nabla_{\rm ad}
\end{cases}\\
\nabla_{\rm rad} &\equiv \dfrac{3 \kappa \rho H_p F_{\rm tot}}{16 \sigma T^4},
\end{align}
where $\nabla_{\rm rad}$ and $\nabla_{\rm ad}$ are the
standard radiative and adiabatic temperature gradients, respectively,
$\nabla_{\rm conv}$ is the convective temperature gradient
computed using mixing length theory \citep{P69,H98},
$\kappa$ is the Rosseland mean opacity, $H_P$ is the pressure scale height, and $F_{\rm tot}$ is the total flux passing through a given height.
Through trial and error we determined that using a value of six for
the mixing length parameter ($\alpha_{\rm ml}=6$) produced vertical
structure models which closely resemble the time and horizontally averaged
vertical profiles measured in the MRI simulations that exhibit
convection (see Figure \ref{fig:vert-mod}).
This is compared
to the much lower and more conventional value of 1.5 used in the DIM by
\citet{H98}.  We note that if the mixing length theory is to
be taken at face value then $\alpha_{\rm ml}=6$ implies that
the length scale of convective eddies is several ($\sim 6$) times larger
than the pressure scale height of the disc, which seems unphysical.

However, it is important to note that this value of the mixing length
parameter actually reflects the fact that when convection occurs in
the simulations, it does so {\it intermittently}.
This intermittency is the result of a limit cycle, operating on timescales of $\sim 10$ thermal times, which is driven by the interplay of temperature dependent opacities and enhancement of stress by convective turbulence \citep[see Section 3.4 of][for further discussion]{HIR14}.
Averaging over this time dependent cycle results in a high effective \aML,
but the time dependent \aML tend to have more canonical values of $\sim 1$ when convection is occurring.
By measuring the horizontally-averaged convective heat flux $F_{\rm conv}$
directly from the simulations, we compute the mixing length parameter
$\alpha_{\rm ml}$ that would produce this flux as a function of
height and time.
We accomplished this
by solving the following equations
using the Newton-Raphson method:
\begin{equation}
\label{eqn:alphaml}
\alpha_{\rm ml}^2\beta^{3/2}=\dfrac{2F_{\rm conv}}{C_P\rho u T},
\end{equation}
where
\begin{align}
u\equiv\sqrt{-\dfrac{g_zH_P}{8}\left(\pder{\ln \rho}{\ln T}\right)_P},\\
H_P\equiv -\text{sign}(z)\pder{z}{\ln P},
\end{align}
$\beta\equiv \left(\nabla - \nabla^\prime\right)$ is the positive root of
the quadratic
\begin{equation}
\beta=\left(\gamma_0\alpha u\right)^2\left(\nabla-\nabla_{\rm ad}-
\beta\right)^2,
\end{equation}
and
\begin{equation}
\label{eqn:gamma0}
\gamma_0\equiv\dfrac{C_P\rho }{8\sigma T^3 \theta},
\;\theta\equiv\dfrac{3\tau}{3+\tau^2},
\;\tau\equiv\alpha_{\rm ml}\kappa\rho H_P.
\end{equation}
Note that $\beta$ and $\gamma_0$ are implicit functions of $\alpha_{\rm ml}$.
$F_{\rm conv}$, $C_P$, $\rho$, $T$, $g_z=-\Omega^2z$, $H_P$, $\nabla$, $\nabla_{\rm ad}$, and
$\kappa$ are read or computed from the MHD simulations. These variables
are then horizontally averaged and smoothed by 1 orbit in time before
$\alpha_{\rm ml}$ is computed.

\begin{figure}
\centering
\includegraphics[width=\linewidth]{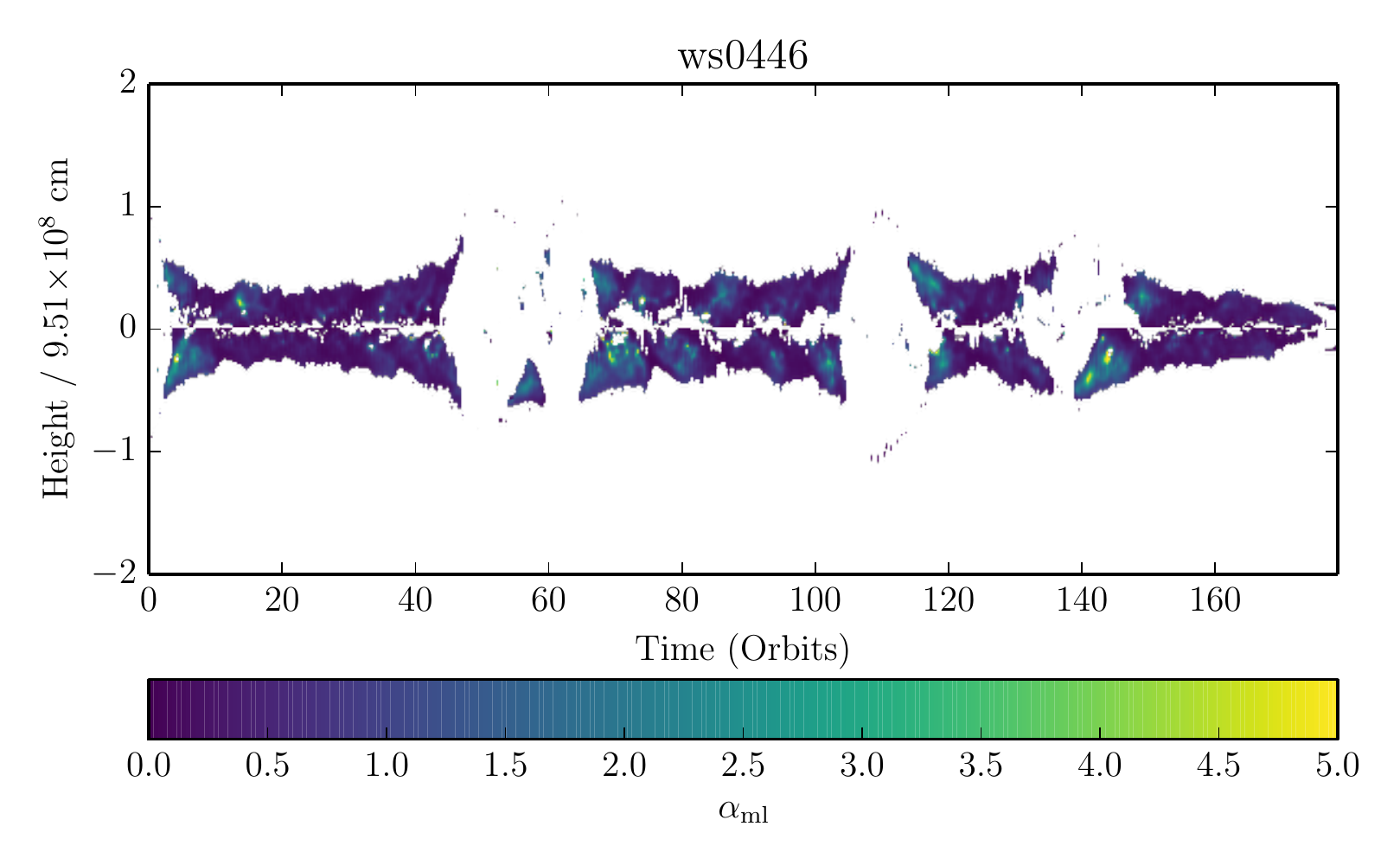}
\caption{
Convective mixing length parameter $\alpha_{\rm ml}$ computed
locally from horizontally-averaged simulation data for upper
branch simulation ws0446 from \citet{HIR14}, as a function of height
and time.  The $\alpha_{\rm ml}$ data (already computed from quantities
smoothed over 1 orbit) has been smoothed by an additional 0.2 orbits in time
to improve clarity.
White regions within two pressure scale heights are either convectively stable,
or are regions for which
our Newton-Raphson method to solve equations (\ref{eqn:alphaml})-(\ref{eqn:gamma0})
failed to converge within 100 iterations.
Data outside two pressure scale heights are discarded.
The white gaps approximately centered
on the 55 orbit, 110 orbit, and 140 orbit epochs are times when radiative
diffusion dominated convection in the vertical transport of heat.
}
\label{fig:alphaml}
\end{figure}

The results are shown in Figure~\ref{fig:alphaml}.  The epochs that are white
at all heights in this figure are epochs when radiative diffusion dominates, and
there is little vertical convective transport of heat.  Most of the epochs
that are actually convective have very reasonable values of $\alpha_{\rm ml}$
that are of order unity.  It is generally only near the epochs that are
radiative that $\alpha_{\rm ml}$ takes on substantially larger values.

The vertical profiles of the simulation data shown in
Figure~\ref{fig:vert-mod} average over both the convective and radiative
epochs, so it is not surprising, given the behavior shown in
Figure~\ref{fig:alphaml}, that unusually large mixing length parameters
are required to describe these profiles with a pure convective transport
treatment.

From the discussion here it is clear that time averaging the intermittent convection obscures some of the physics discovered in the MRI simulations of \citet{HIR14}. It is unclear how much this simplification affects the outcome. The duration of the convective limit cycle is $\sim 50$ orbits, so for timescales $\gtrsim 1$ day, this averaging procedure should be a reasonable approximation. It is during the rapid transition from quiescence to outburst that this simplification becomes questionable, making the inability of the DIM to capture this time dependent behavior a clear limitation, but nothing better can be done in the framework of the standard DIM.

\subsection{Summary of MRI Simulation-Based Vertical Structure Equations and Boundary Conditions}
\label{sec:SumEqn}
The DIM framework uses the thermal pressure to provide both the vertical
hydrostatic support and to specify the vertical dissipation profile.
In the simulations, magnetic pressure support can dominate thermal pressure
near the photosphere, although the magnetic to thermal pressure ratio
is at most $1.5\alpha \lesssim 20\%$ near the midplane.
Including magnetic pressure in a single pressure framework is complicated,
as it requires a modification of the alpha relation
(equation \ref{eqn:SSalpha}) and it complicates the temperature gradient
(equation \ref{eqn:T-grad}) used in mixing length theory.  We have
therefore neglected this additional aspect of the simulation physics.
Additionally, the DIM uses different, albeit similar,  equation of state and opacity tables.

With our modifications to the DIM our vertical structure equations become
\begin{align}
\label{eqn:HSE}
\der{P}{z}&=-\rho\Omega^2z\\
\der{\varsigma}{z}&=2\rho\\
\der{\ln T}{\ln P}&=\nabla\\
\label{eqn:flux}
\der{F_z}{z}=Q^+&=A_0\Omega P_{0}\left(\dfrac{P}{P_0}\right)^\delta\\
\label{eqn:flux-norm}
A_0&=\dfrac{3}{2}\alpha\dfrac{P_{0}^{\delta -1}\int_0^\infty P\dd\! z}
{\int_0^\infty P^\delta\dd\! z},
\end{align}
where $\varsigma (z)$ is the surface mass density between $\pm z$, and $\nabla$ is determined by equation (\ref{eqn:T-grad}).
Equations (\ref{eqn:flux}) and
(\ref{eqn:flux-norm}) are equivalent to our dissipation fit and the alpha
relation (equations \ref{eqn:Qfit} and \ref{eqn:SSalpha}, respectively).
Our midplane boundary conditions are
\begin{align}
z&=0\\
F_z&=0\\
\varsigma&=0\\
T&=T_0\\
P&=P_{0},
\end{align}
and our exterior boundary conditions are
\begin{align}
\kappa_R P&=\dfrac{2}{3}\Omega^2 z\\
F_z&=\sigma T^4\\
\varsigma&=\Sigma,
\label{eqn:mrivertstruc2}
\end{align}
where $\kappa_R$ is the Rosseland mean opacity.

\subsection{Thermal equilibria: the S-curves}

\begin{figure}
\centering
\includegraphics[width=\linewidth]{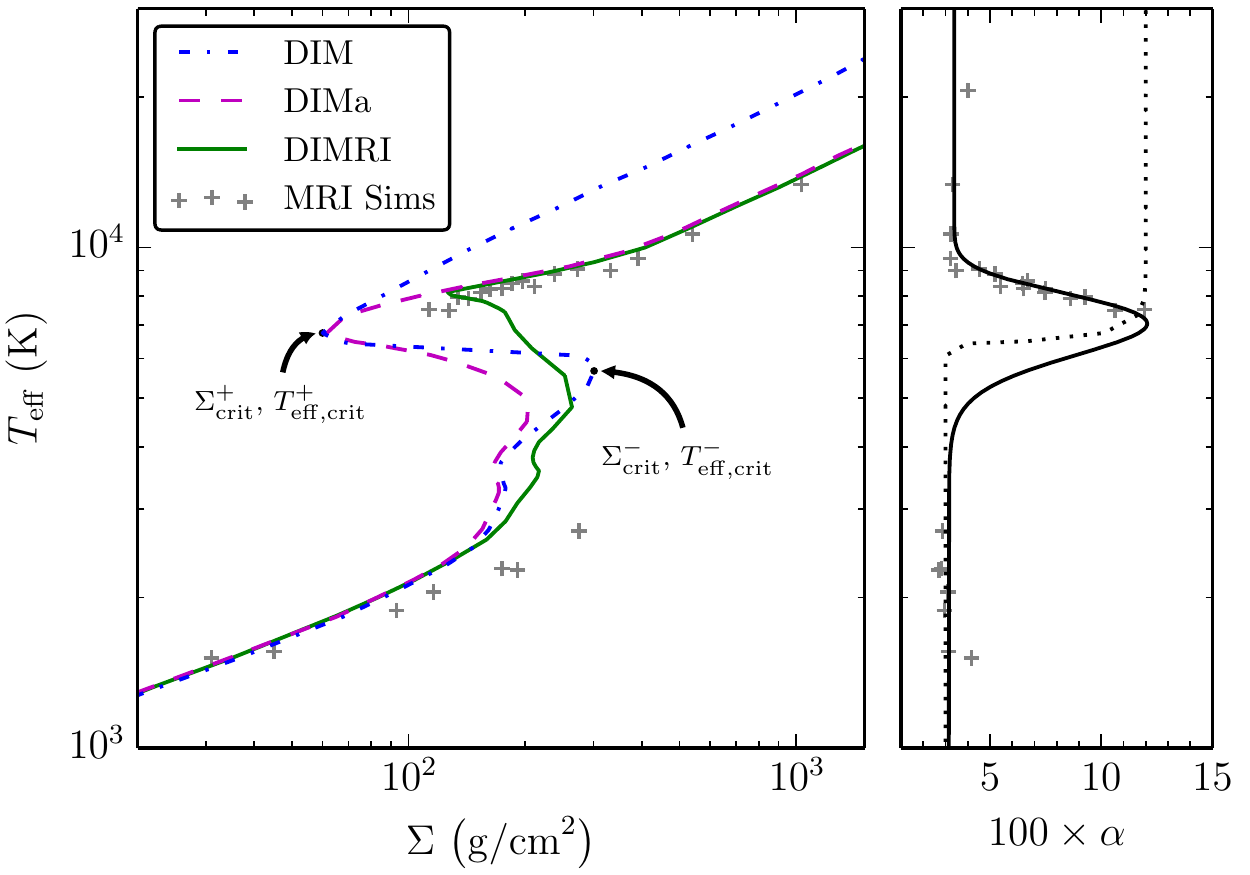}
\caption{
Left: Loci of thermal equilibria in the $T_{\rm eff}$ vs. surface mass density
$\Sigma$ plane (the ``S-curve") at radius $R=1.25\times10^{10}$ cm for the
standard DIM (blue), DIMa (magenta) and DIMRI (green). The MRI simulation results are gray crosses (one point for each stable simulation). Additionally, the critical points (\Scp,~\Teffp) and (\Scm,~\Teffm) are marked for DIM.
Right: The MRI-based $\alpha\left( T_{\rm eff} \right)$ fit from
Figure~\ref{fig:aT} plotted sideways in solid black (DIMa and DIMRI both use this fit) with the results from the same
MRI simulations from the left plotted as gray crosses.
Equation (\ref{eqn:alphadim}) for the standard DIM, with $\ac=0.03$ and
$\ah=0.12$, is plotted as the dotted black line.
}
\label{fig:s-curve}
\end{figure}

Before presenting outburst light curves based on the physical
models discussed in the previous two sections we briefly discuss the
properties of the disc's thermal equilibria.  We consider two
MRI-based models, DIMa and DIMRI, and compare them to the standard DIM.
DIMa adopts the same vertical structure assumptions as the standard
DIM, but uses the MRI-based $\alpha(T_{\rm eff})$ prescription
of equations (\ref{eqn:alphamri1})-(\ref{eqn:alphamri2}) discussed
in section 2. DIMRI also uses this MRI-based $\alpha$ prescription,
combined with the MRI-based vertical structure equations
(\ref{eqn:HSE})-(\ref{eqn:mrivertstruc2}) summarized in section \ref{sec:SumEqn}.
Figure~\ref{fig:s-curve} illustrates the differences in the S-curves
produced by these models at radius $R=1.25\times10^{10}$ cm around a
$0.6M_\odot$ white dwarf.

The most important parameters emerging from the S-curves shown
in Figure~\ref{fig:s-curve} are the critical surface mass densities
(\Scp, \Scm) and effective temperatures (\Teffp, \Teffm) at the ends
of the (upper, lower)
branches, marking the points where (cooling, heating) transitions occur.
In particular, the quotient of critical surface mass densities, $\QSc\equiv\Scm/\Scp$, plays an important role
in determining the shape of the outburst lightcurve.
It is therefore worth noting that \QSc is different for all of the S-curves,
with DIM having the largest \QSc.
Despite our efforts, there is still a basic discrepancy between
the actual MRI simulation data
and the DIMRI S-curve in Figure \ref{fig:s-curve}.
Both branches of the simulation S-curve extend a little further in $\Sigma$, leading to a larger \QSc compared to that of DIMRI,
and \Teffm is significantly larger in DIMRI on the lower branch.
The mismatch on the upper
branch can be explained by our choice of $\alpha_{\rm ml}=6$.
As an annulus in outburst approaches the critical point (\Scp, \Teffp) the role of convection increases and presumably $\alpha_{\rm ml}=6$ becomes increasingly less adequate. This is because the high value of \aML that we adopt is due to radiation dominated episodes in the intermittent convection, which become less prominent as the annulus reaches the end of the upper branch. One possible solution would be to decrease \aML towards the tip of the upper branch in the DIMRI,
and this may be worth exploring in the future.  We show how adopting a smaller constant
value of \aML in DIMRI affects the outburst lightcurves below.

There are several issues which contribute to the mismatch between the
simulation data and DIMRI on the lower branch, and these all stem from
the fact that we have been unable to find stable thermal equilibria in
the simulations for effective temperatures higher than 3000~K.  At such
low temperatures, the opacities are so small that these
equilibria are only marginally optically thick,
with midplane optical depths $\tau\lesssim 5$.  This may be a problem given
that the simulations assume flux limited diffusion which may not accurately
treat radiation transport at such low optical depths.
The corresponding DIM models (i.e. DIM, DIMa, DIMRI) at these low temperatures
also have low optical depths.  As a consequence, the density at the
photosphere (the exterior boundary condition) is a significant fraction of
the central density implying that a large fraction of mass ($\sim 10-50 \%$)
is ignored/neglected\footnote{
These inaccuracies do not affect standard DIM models of real
outbursts, because in these models the disc never cools down to
temperatures at which these discrepancies appear.  In the standard DIM one
simply tunes \ah and \ac to obtain the required ratio of the critical surface
densities.
}.
This missing mass likely has a significant impact on
the temperature and density profiles of DIM models and is, at least partly,
responsible for the miss-match between the DIMRI curve and the simulation
data on the lower branch of the S-curve. The influence that this missing
mass has on the temperature profiles may also explain why the DIMRI models
near the end of the lower branch are convective (which is why the DIMRI
S-curve turns up at $\Sigma\approx180$ g cm$^{-2}$ in
Figure \ref{fig:s-curve}), while there is no convection in any of the lower
branch MRI simulations.   Again, we have been unable to find any stable
thermal equilibria at higher temperatures on the lower branch, where the
opacities would be higher and missing mass would be less of an issue in the
DIM models.  However, there is perhaps an even bigger problem with the
simulations on the lower branch, and that is that they neglect non-ideal
MHD effects.  As we demonstrate in Appendix \ref{sec:non-ideal}, Ohmic
resistivity and the Hall term are likely to be important here, and so this also
casts uncertainty on the lower branch results, and specifically the critical point of the lower branch (\Scm, \Tcm) which contributes to the determination of \QSc.
Future work will have to account
for these effects on the lower branch.  Our focus here, however, is to see
whether the variation in $\alpha$ on the upper branch that has been found
in the simulations can produce reasonable dwarf nova light curves.

Coming to the differences between our variants of the DIM, we first
note that \Scm and \Teffm at the end of the lower branch are much higher
on the classic DIM S-curve compared to both the DIMa and DIMRI S-curves,
consequently DIM has the largest \QSc.
As DIMa shares exactly the same vertical structure assumptions and equations
as the classic DIM, this can only be due to the different alpha-prescriptions
between the two models (equations \ref{eqn:alphamri1}-\ref{eqn:alphamri2}
and equation \ref{eqn:alphadim}, respectively).
Recall from
Figure~\ref{fig:aT} that our fit to the simulation data has $\alpha$ starting
to increase at $T_{\rm eff}\sim4000$~K, roughly indicating the end of the lower branch.
By contrast, the choice of $T_{\rm crit}^+$ in the classic DIM
equation
(\ref{eqn:alphadim}) corresponds to a much higher effective temperature
($\sim6000$~K) at the end of the lower branch.

It should be noted that we have no simulation data for
3000~K$\lesssim T_{\rm eff}\lesssim$7000~K in Figure \ref{fig:aT}, precisely because the simulations failed
to produce stable thermal equilibria in this range.
Therefore, our fit in equations \ref{eqn:alphamri1}-\ref{eqn:alphamri2} has
some flexibility in this temperature range.
In principle, we could have
fit the simulation data in Figure~\ref{fig:aT} with a function that keeps
$\alpha$ low until the effective temperature increases above 6000~K,
and that would bring the lower branches of the DIMa and DIM S-curves into much
better agreement.
However, simulation ws0467 shown
in Figure~\ref{fig:ws0467}, was started near the end of the
lower branch at $T_{\rm eff}\simeq3000$~K and underwent runaway heating\footnote{
Although this helps constrain the end of the lower branch, it is numerically challenging to further resolve the critical point (\Scm, \Tcm), as the simple act of relaxing from the initial conditions could push a simulation too far from (\Scm, \Tcm) to obtain a thermal equilibrium.}.
Hence our fits (equations \ref{eqn:alphamri1}-\ref{eqn:alphamri2}) produce
S-curves that better represent the behavior observed in the simulations.

Another alternative to achieve agreement on the lower branch is to reduce
$T_{\mathrm{c0}}$ in the classic DIM.
This has actually already been done by \citet{H02}, who modified
this parameter to 8000~K, thereby producing a lower branch that only
extended up to an effective temperature of 3000~K.  This modification
produces very similar outburst light curves, except that the quiescent light
curves are flatter in shape, which actually may agree somewhat better with
observations \citep{H02}.
Whether
the low \Teffm
can be claimed as a success of
the MRI simulations will require a full treatment of the non-ideal MHD
effects that have so far been neglected, but are likely to be crucial
in the quiescent state,
and hence will likely shift the end of the lower branch.

As we discuss in more detail below, the new simulation-based vertical
structure equations (most importantly the different \aML) are the reason
DIMRI has a different location for the end of the upper branch than DIM and DIMa in Figure \ref{fig:s-curve}.
In fact, if we plot a DIMRI S-curve with $\alpha_{\rm ml}=1.5$, the end of the
upper branch matches that of the DIM and DIMa S-curves.  The very large
effective mixing length parameter $\aML=6$ used
in DIMRI results in more efficient convective transport, which flattens
the temperature profile thus
increasing \Teff for a fixed midplane temperature \Tc.  Because the opacity
is largely determined by \Tc, it is \Tc which determines where the end
of the upper branch occurs. For radiative cooling the effective and central temperatures
are related through
\begin{equation}
\Tc=\left( \frac{3\tau_{\rm tot}}{8} \right)^{1/4}\Teff,
\label{eq:tcteff}
\end{equation}
where $\tau_{\rm tot}$ is the total (vertical) disc opacity \citep[see e.g.][]{DUB99, Kotkoetal12}.
For a fully ionized disc the Rosseland opacity is
$\kappa \sim \Sigma H^{-1}\Tc^{-7/2} \mathrm{cm^2/g}$ from which follows the well known \citep{SS73}
relation $\Teff \propto \Sigma^{5/14}$. Recombination breaks this relation and changes
its slope while convection enhances cooling.  The result is that the upper
branch ends at higher effective temperature, and hence higher surface density.

\section{Outburst Light Curves}

\subsection{The outburst cycle}

During the low luminosity ``quiescent" phase of a dwarf nova cycle,
the effective temperature in the whole disc is lower than \Teffm
(the disc is
cold). The mass accretion rate is not constant across the cold disc so the
disc accumulates matter, increasing its temperature and surface density (each disc
annulus moves up the lower branch of its local S-curve).
Finally, at some radius, $R_0$, the accumulation time becomes shorter than the viscous time, and the ionization of gas becomes so significant that the local cooling mechanisms become inefficient and thermal equilibrium is lost.
The annulus at $R_0$ undergoes rapid heating and makes the transition to the hot state (upper branch of its local S-curve).
A narrow fully-ionized region of high viscosity has just formed at $R_0$, and is surrounded by cold matter.
This induces a steep radial temperature gradient and the formation of the
heating front, indicating the beginning of an outburst.
Additionally, a spike in the \Sig profile arises as a consequence of different viscous efficiencies in the hot and cold parts of the disc: the low viscosity outside the hot annulus provides insufficient outwards angular momentum transport to prevent further accumulation of mass at $R_0$.
These steep \Sig and \Teff gradients in the heating front cause matter and
heat to diffuse to the adjacent annuli, forcing their transition to the hot
state. The hot region in the disc widens as the heating front propagates
through the disc causing the luminosity to rise. The elevated mass accretion
rate in the hot region reduces the surface density behind the heating front
and enhances the mass inflow to the inner disc region\footnote{We describe here
an {\sl inside-out} outburst. For high mass-transfer rates, outbursts can be of  {\sl outside-in}
type, i.e. the heating front propagates inwards from the disc outer regions \citep[see e.g.][]{L01}.}.

Once the heating front reaches the outer disc edge, the disc is fully ionized
and reaches its maximum brightness (the outburst maximum).
In the DIM, the minimum critical surface density $\Scp$ at the end
of the upper branch of the local S-curve is approximately
proportional to radius $R$ \citep{H98}, causing $\Scp$ to be highest at
$\Rout$.
Consequently \Sig manages to only rise slightly above this critical
value near $\Rout$
as the heating front passes and it falls below the critical value
almost immediately after the front dissipates.
At the radius where this
happens the cooling is strongly enhanced by the change of opacities when
\Teff$<$\Teffp and \Sig and \Teff gradients lead to the formation of the
cooling front.
The inward propagation of the cooling front through the disc leads to the
observed outburst decay.

The cooling front develops at the outer disc edge almost at the same time as the heating front disappears, allowing no time for the mass accumulated in the outer parts of the disc to arrive at the inner disc radius before the cooling front sets in,
even though the mass accretion
rate everywhere in the hot disc has increased beyond the mass transfer rate from
the secondary.
The surface density profile at the outburst maximum is not yet proportional to $R^{-3/4}$ as expected in a hot stable disc.
Therefore, even after the development of the cooling front, the mass accretion rate near the inner disc edge still increases until the mass excess from the outer disc region has traveled through the whole disc and has been accreted onto the white dwarf (see Figure \ref{fig:fronts} and for more details \citealt{K12}).
Only after this will the hot region ahead of the cooling front reach the hot stable state where the constant mass accretion rate is of order \Mtr, the mass
transfer rate from the secondary.

\begin{figure*}
\includegraphics[width=\linewidth]{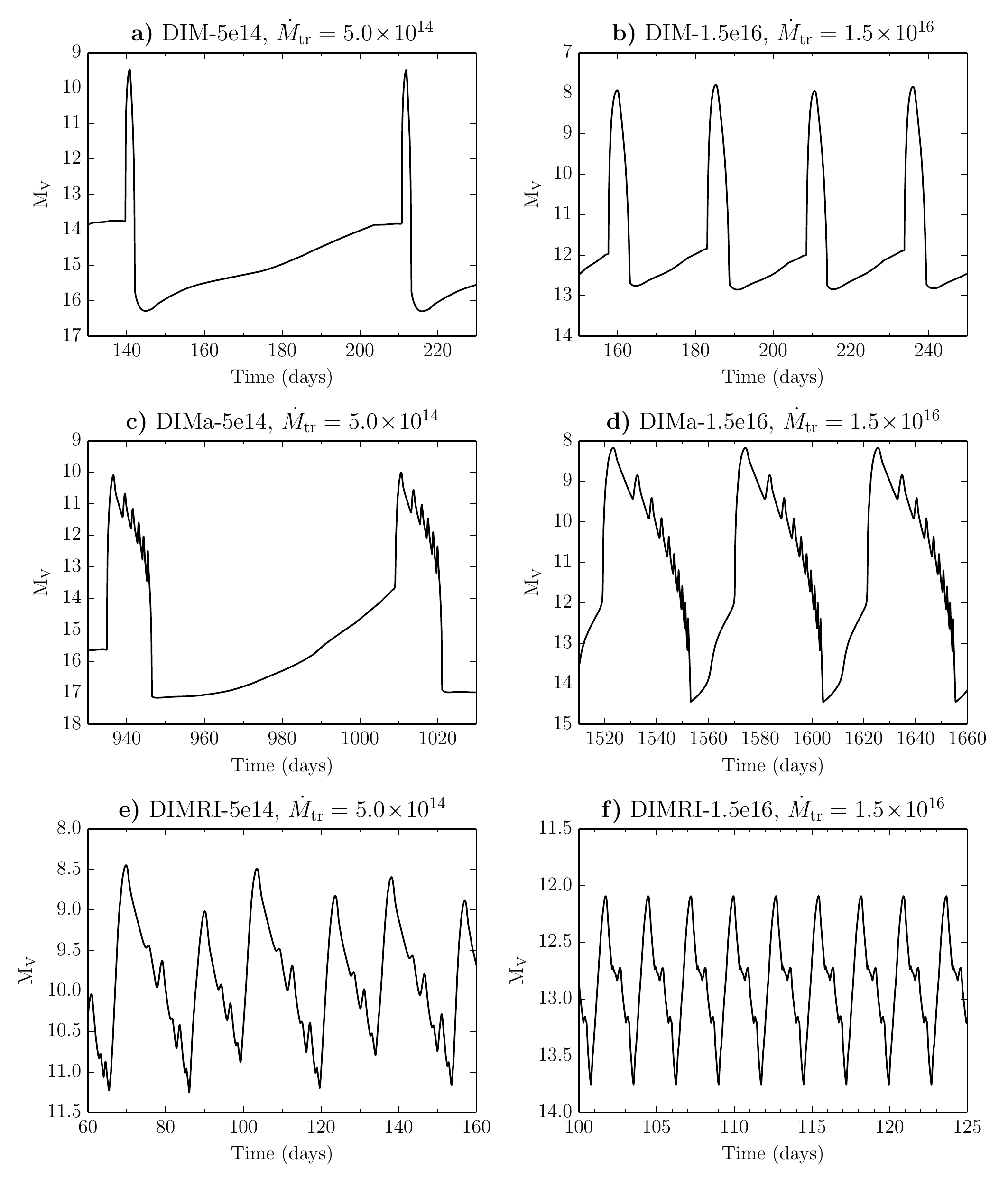}
\caption{Visual magnitude lightcurves for six of the models listed in Table \ref{Tab:LC_param}.
The sawtooth-like features are reflares.
The units for the listed mass transfer rates are g/s.
}
\label{fig:lightcurves}
\end{figure*}

\begin{table*}
\begin{tabular}{|l|c|c|c|c|c|}
\hline
Model & \Mtr [g/s] & $A_{\rm outb}$ & $T_{\rm outb}$ & $T_{\rm quiesc}$ & Figure \tabularnewline
\hline
\hline
DIM-5e14 & $5\times10^{14}$ & $5.4$ & $2.3$ & $47.3$ & \ref{fig:lightcurves}a \tabularnewline
\hline
DIM-1.5e16 & $1.5\times10^{16}$ & $4.6$ & $6.3$ & $11.6$ & \ref{fig:lightcurves}b \tabularnewline
\hline
DIMa-5e14& $5\times10^{14}$ & $7$ & $11.6$ & $46.4$ & \ref{fig:lightcurves}c \tabularnewline
\hline
DIMa-1.5e16& $1.5\times10^{16}$ & $6.3$ & $50$ & $0$ & \ref{fig:lightcurves}d \tabularnewline
\hline
DIMRI-5e14 & $5\times10^{14}$ & $1.7$ & $2.8$ & $0$ & \ref{fig:lightcurves}e \tabularnewline
\hline
DIMRI-1.5e16 & $1.5\times10^{16}$ & $2.6$ & $19.9$ & $0$ & \ref{fig:lightcurves}f \tabularnewline
\hline
DIMRI-$\alpha_{ML}=1.5$ & $5\times10^{14}$ & $7.7$ & $17.3$ & $60.1$ & \ref{fig:aml} \tabularnewline
\hline
DIMRI-6e13 & $6\times10^{13}$ & $3.2$ & $1.9$ & $7$ & \ref{fig:DIMRI-6e13} \tabularnewline
\hline
DIMRItr-1.5e16 & $1.5\times10^{16}$ & $3.4$ & $14.6$ & $4.7$ & \ref{fig:amltrunc} \tabularnewline
\hline
\end{tabular}
\caption{The parameters of the outbursts measured in our calculated
lightcurves. $A_{\rm outb}$ is the outburst amplitude in magnitudes,
$T_{\rm outb}$ is the outburst duration time in days and $T_{\rm quiesc}$ is the quiescence time in days, i.e. the time elapsed between the end of an outburst and the beginning of the next. The last column lists the figure where the
lightcurve for a given model can be found.
}
\label{Tab:LC_param}
\end{table*}

\subsection{Reflares}

Due to the high \Mtr and
low \ac, the disc may accumulate a lot of mass
during the quiescent phase and rise to an outburst.
If the viscosity in the hot disc is not efficient enough (i.e. when \ah is relatively low) to redistribute the mass excess accumulated during the previous outburst phases to the inner region (where it can be accreted),
or if the critical points $(\Scm,\Tcm)$ and $(\Scp,\Tcp)$ are too close (i.e. \QSc is too small),
the fronts propagating in the disc (both cooling and heating) may be stopped before arriving at either of the two disc edges.
This gives rise to the appearance of reflares in the outburst lightcurves.  As we discuss
below, all our MRI-based models exhibit this phenomenon, and we therefore begin with a
brief description of the cause: it is the confluence of the
small \ah high on the upper branch as well as the small \QSc that is responsible for these reflares.

As a cooling front propagates inward, the high viscosity of the
hot matter inside the cooling front contrasted with the low viscosity outside
the cooling front causes an outward diffusion of matter across the front.
This in turn causes a deficit in surface density within the front itself,
followed by an enhanced surface density in the cold region behind (outside)
the front.  Hence, as the cooling
front moves inward in radius, at some point the post-front \Sig may become
high enough to cross the critical value \Scm at the end
of the lower branch of the local S-curve.
It is here where it is clearest that the low \ah and the lack of sufficient separation between \Scm and \Scp (i.e. too small \QSc) 
conspire against the smooth propagation of a cooling front by creating a mass excess outside the front and setting a low critical threshold respectively.

In this situation a new heating front
arises and starts to move outwards. The matter heated by this newly formed
front flows at a high rate into the zone of the cooling front, increasing its
temperature and surface density. This inflow of hot gas eventually destroys
the cooling front and only a heating front is left. As a result, the inward
propagating cooling front behaves as if it is reflected into an outward
propagating heating front  before it arrives at the inner edge.
A similar mechanism can cause reflection of the heating front propagating
toward the outer disc radius. If the post-front \Sig remains close to \Scp the
elevated accretion rate in the hot region behind the front may cause \Sig to
fall below the critical value and a cooling front will start to form. The
reduced transport of the angular momentum through the emerging cold zone
finally stops the propagation of the heating front and a newly formed
cooling front will move inward.  These reflections produce a reflare pattern
in the outburst lightcurves
(see sawtooth-like features in Figure \ref{fig:lightcurves}),
which are not observed in standard dwarf novae.
As reviewed in section 4.3 of \citet{L01},
reflares are a common feature of the DIM, and one must work to get rid of them
by choosing appropriate values of $\ah$ and $\ac$ in order to agree with the
smooth observed light curves. As we will see shortly, reflares are
also a generic feature of all our MRI-based lightcurves.
However, it is important to reemphasize that the reflares (and the details of outbursts) are dependent on where the lower branch ends; a detail we do not claim to model accurately.

\subsection{Results}

The outburst properties depend on disc viscosity and the parameters characterizing
a binary.
For the new DIMRI to be considered as a possible replacement for the standard DIM,
it should reproduce the basic features of dwarf novae
lightcurves such as outburst amplitude, outburst duration and quiescence duration.

To better understand how the new $\alpha$-prescription and the new disc
vertical structures introduced into the classical DIM influence outburst
light curves, we calculated these
light curves using three models: DIMRI, DIM with $\alpha$ as a function of \Teff
(DIMa) and classic DIM with $\ac=0.03$ and $\ah=0.12$.
All models were calculated for the same set of parameters: a primary mass $M_1=0.6\, \Msun$, the disc inner radius $R_{\rm in}=8.67\times10^8$ cm,
and the circularization radius $R_{\rm circ}=2.85\times10^{10}$ cm.
In addition we run the calculation for two different values of the mass transfer rate: $\Mtr=1.5\times10^{16}$ g/s and $\Mtr=5\times10^{14}$ g/s for each model (see Table \ref{Tab:LC_param}).
In all models the outer disc radius is variable due to the fact that we take into account the tidal force acting between the secondary star and the disc.
For the models in this paper, the average outer disc radius is
$\left\langle R_{\rm out} \right\rangle=4.6\times10^{10}$ cm.

\begin{figure}
\includegraphics[width=\linewidth]{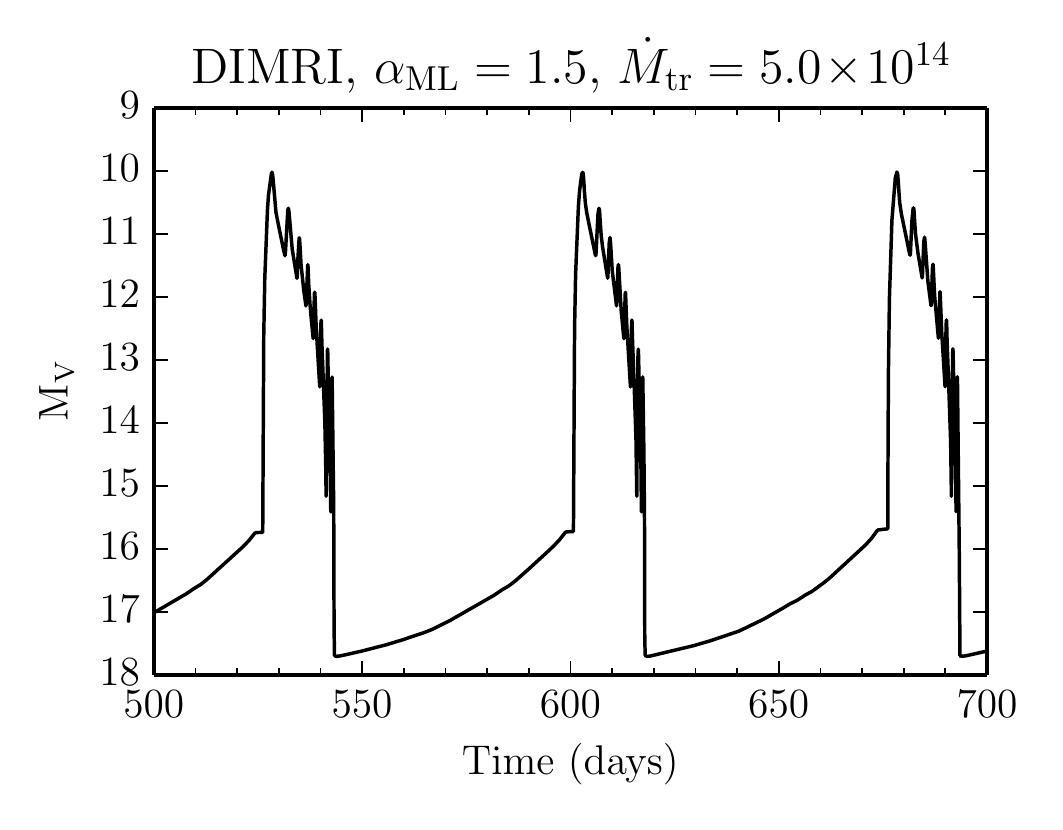}
\caption{Light curves calculated from DIMRI but for \aML$=1.5$; $\dot M_{\rm tr}=5.0 \times 10^{14}\,\rm g\,s^{-1}$. Compare with Fig.~\ref{fig:lightcurves}e.}
\label{fig:aml}
\end{figure}

\begin{figure}
\includegraphics[width=\linewidth]{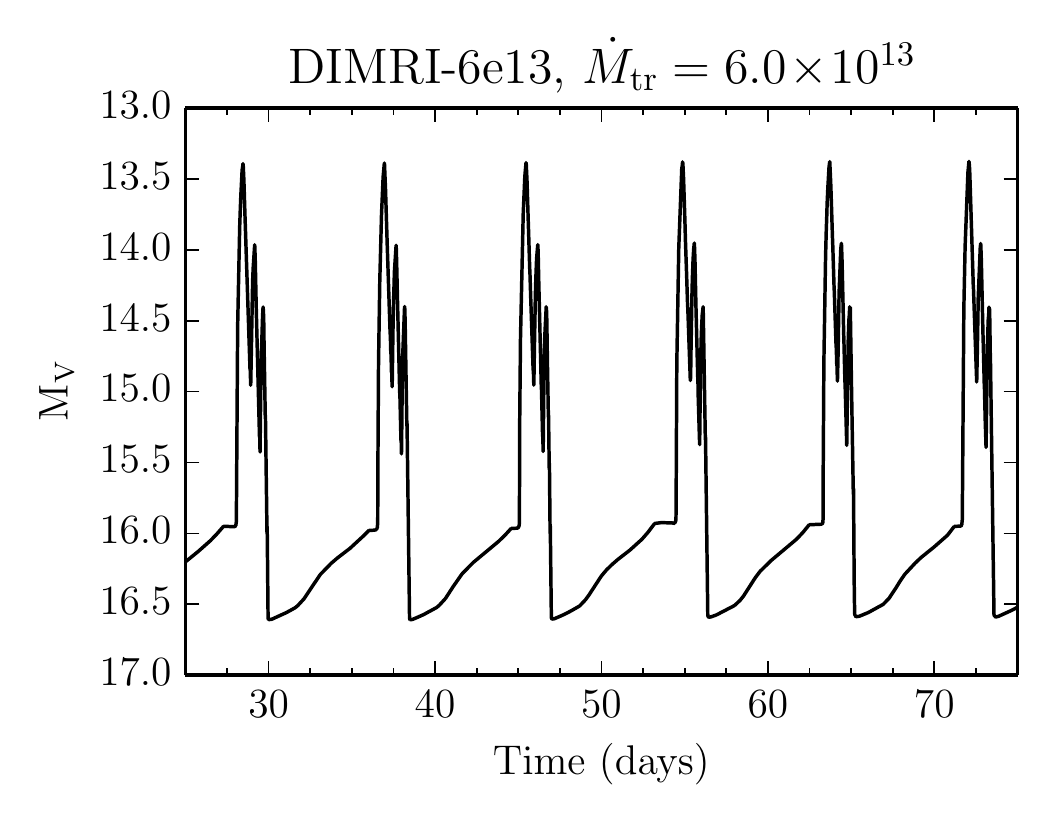}
\caption{Light curves calculated from DIMRI with $\dot M_{\rm tr}=6.0 \times 10^{13}\,\rm g\,s^{-1}$.}
\label{fig:DIMRI-6e13}
\end{figure}

\begin{figure}
\includegraphics[width=\linewidth]{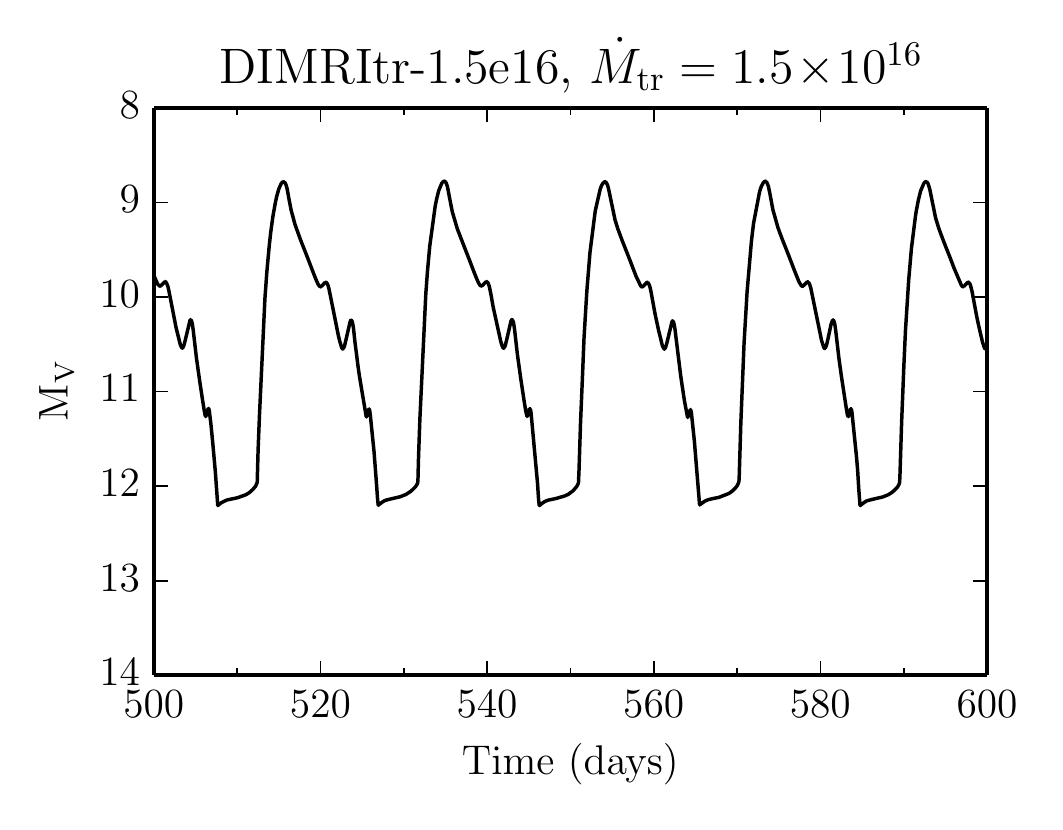}
\caption{Light curves calculated from DIMRI with inner disc radius truncated by the magnetic field with magnetic moment $\mu=8\times 10^{30} $G cm$^3$ for \aML$=6.0$; $\dot M_{\rm tr}=1.5 \times 10^{16}\,\rm g\,s^{-1}$.
As can be seen by comparing this to Fig.~\ref{fig:lightcurves}f, the
truncation of the inner disc results in the appearance of quiescence and more
regular outbursts.
}
\label{fig:amltrunc}
\end{figure}

Analyzing the differences between the lightcurves
(see Figures \ref{fig:lightcurves}-\ref{fig:amltrunc})
gives insight into the physical implications of our modifications to the DIM.
One subtle difference between the classic DIM lightcurves and the MRI based lightcurves is that the MRI based ones are not strictly periodic.
We do not understand why the new $\alpha$ prescription causes this.
One possibility is
that the disc needs much more time to relax with this prescription.
Conversely,
the most striking difference between the DIM and the two other model light curves is that the outburst decay in DIM is smooth while in DIMRI and DIMa the outburst decay has small amplitude brightness variations
characteristic of reflares (compare Figure~\ref{fig:lightcurves}c-f with Figure~\ref{fig:lightcurves}a,b). The reason that reflares do not appear in the DIM but are present in the two other models
is connected to our $\alpha$-prescription, but is also tied to our uncertainties on the lower branch, specifically the location of the critical point (\Scm, \Tcm).
In the DIM, 
\QSc is larger than the other models and
$\alpha$ maintains its high value ($\ah=0.12 \gg \ac$) in the whole hot part of the disc until the cooling front passage. In the models where $\alpha$ is a function of effective temperature, the higher viscosity is present in a much more narrow region, i.e. in which the central temperature is higher than \Tcm but lower than $5\times10^4 {\rm K}$.
Therefore, the mass in the inner disc region is accreted at a lower rate than in the DIM due to the lowered viscosity in the high
temperature regime,
leading to an excess of mass behind (outside) cooling fronts.
It is the combination this effect coupled with the small \QSc which is directly responsible for the reflares seen in DIMa and DIMRI.
Furthermore, \QSc as determined by MRI simulations is actually larger than that found in DIMRI, which suggests that DIMRI is more susceptible to reflares than what the simulation data imply.

\begin{figure*}
\includegraphics[width=\linewidth]{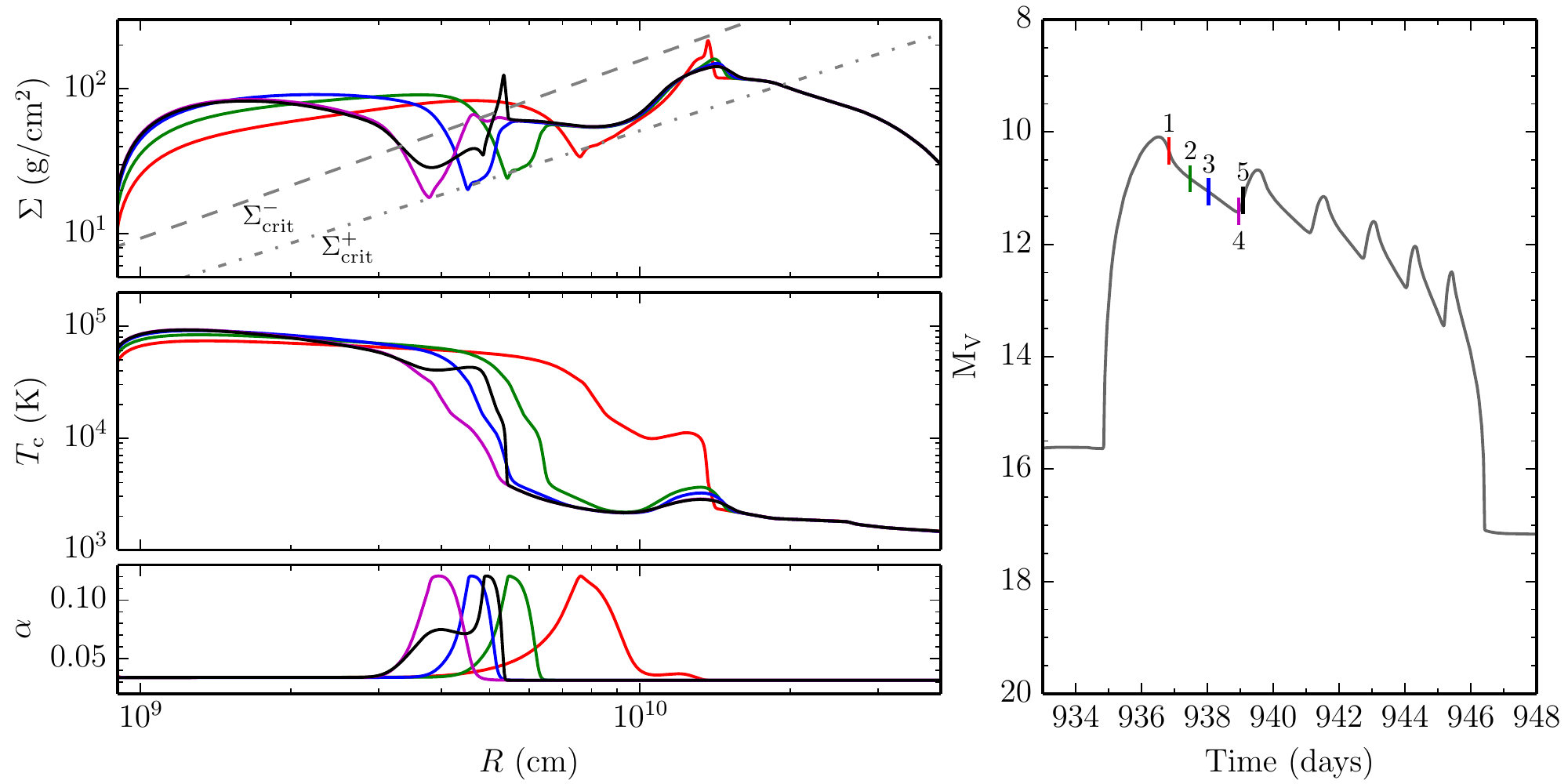}
\caption{Left panels: Radial profiles of surface density (top), midplane temperature
(middle), and $\alpha$-parameter (bottom) during the initial decay from
outburst for the DIMa calculation with $\dot{M}=5\times10^{14}$~g~s$^{-1}$.
Right panel: Zoom of the first outburst in the lightcurve  Figure~\ref{fig:lightcurves}(c).
Different successive times 1-5 are shown by different colors as indicated,
with 1 (red) being the first and 5 (black)
being the last.
It is important to note that $\alpha$ is small for $R<3\times 10^9$~cm for all epochs shown despite
the fact that the disc is in outburst, because we are high on the upper branch here. This results
in the elevated surface mass density that can be seen in the inner disc. As the cooling front (the dip/rarefaction in $\Sigma$ located at $R\approx7.6\times 10^9$~cm in epoch 1) propagates inwards through the disc, mass is redistributed from ahead of the front where $\Sigma$ is high to the post-front region. Eventually, the post-rarefaction surface density
crosses the critical surface density $\Sigma_{\rm crit}^-$ at the end of the
lower branch of the local S-curve. This occurs approximately at epoch 4 (magenta), which is when the inward propagating cooling front is reflected into the heating front seen at $R\approx5.3\times10^9$~cm in epoch 5 (black). This figure clearly shows that either further separating \Scm and \Scp (i.e. increasing \QSc) or reducing the excess mass in the inner disc by increasing $\alpha$ high on the
upper branch could help alleviate reflares.
}
\label{fig:fronts}
\end{figure*}

Figure~\ref{fig:fronts} shows how the higher surface mass density in the
inner disc and smaller \QSc in DIMa leads to reflections of the inward propagating cooling
front and reflares.  The first epoch shown (the red curves labeled 1)
corresponds to the time when the outward propagating heating front has just
arrived in the outer disc, which is why there is a spike in surface density
and bump in midplane temperature at $R\sim1.3\times10^{10}$~cm.  By this time an
inward propagating cooling front has already been launched, and is located
at $\simeq7.6\times10^9$~cm where the surface density has reached the critical
surface density $\Sigma_{\rm crit}^+$ on the upper branch of the local S-curve.
As the front propagates inward from 1 (red) to 2 (green) to 3 (blue), the
gradients in viscosity cause outward mass diffusion, thereby producing a
rarefaction in surface density down to $\Sigma_{\rm crit}^+$ within the
cooling front, followed by an enhanced surface density behind (outside)
the front.  Because the inner disc in the outburst state has such a high
surface density due to the {\it low} values of $\alpha$ high up on the
upper branch relative to the DIM, the post-front excess in surface mass
density is also high, and eventually, at epoch 4 (magenta) at
$R=4.6\times10^9$~cm, reaches the
critical surface density $\Sigma_{\rm crit}^-$ at the end of the lower
branch of the local S-curve (highlighting the role of small \QSc).
This triggers a heating front which then
propagates outward, as evident in epoch 5 (black) at $R=5.3\times10^9$~cm.

Hence, directly as a consequence of the small \Scm (or alternatively small \QSc)
combined with the
low $\alpha$'s and
resulting
higher surface
densities in the inner disc in outburst, the cooling front that would normally
cause a transition back to quiescence propagates instead with difficulty,
through a sequence of reflections seen as reflares in the light curve. In
contrast, in the DIM with suitably chosen $\ah$ and $\ac$,
there is a larger \Scm making it harder to trigger a reflare. Additionally,
the mass diffusion and accretion during the outburst is
much higher and the inner disc is able to process sufficient mass to lower
the inner surface density and avoid the appearance of the reflares during the
outburst decay.
The contested propagation of the cooling front in DIMa causes the outburst decay phase to last longer compared to DIM (for example compare Figure~\ref{fig:lightcurves}a with Figure~\ref{fig:lightcurves}c).
During this time more mass is being accreted onto the white dwarf in the DIMa,
leaving the disc less massive and less luminous than in the DIM at the end of outburst.
This results in a higher amplitude outburst for the DIMa,
which highlights the effect reflares have on outbursts.

The difference in light curves between DIMRI and DIMa is due to the different
$\aML$ values and different dissipation profiles.
The mixing length parameter $\aML$ is the most important difference, as can
be seen by comparing Figure~\ref{fig:lightcurves}e with Figure~\ref{fig:aml},
which presents the same DIMRI calculation but with $\aML$ restored to its
traditional value of 1.5.
Convection sets in close to the point where the cold branch of the S-curve
ends which leaves this point sensitive to \aML.
This effect is even stronger on the critical point where the hot branch
starts, as this is where convection is the strongest.
Higher values of \aML shift both critical points at which the S-curve bends
closer together, leading to a smaller \QSc.
This alters the global behavior of the disc and the outburst properties: the
decay from the outburst in a disc with $\aML=6$ starts at higher \Sig and
higher \Teff, and less mass is accreted and accumulated in the disc during
the outburst cycle. Therefore, more efficient convection produces outbursts
that are more frequent and of lower amplitude, and even lack quiescent phases. 
While it is important to note that this lack of quiescence may be related to uncertainties in the end of the lower branch, there are other ways out. In the discussion below, we examine a few options to restore or modify quiescence.

The comparison of DIMRI with $\aML=1.5$ and DIMa (Figure \ref{fig:aml} and
Figure \ref{fig:lightcurves}c) highlights the importance of changing the
dissipation profile, as this is the only difference between these two models.
The main difference is that the outbursts in DIMRI with $\aML=1.5$ are wider
and the quiescence is longer than in the DIMa light curve.

Changing the mixing length parameter is not the only way to increase the
outburst amplitude and restore quiescent phases to the DIMRI models.
Increasing the mass transfer rate, while all other parameters are fixed,
makes the disc hotter and denser.
This means that the surface density everywhere in the quiescent disc is closer
to the critical value and the disc luminosity is higher.
The result is more frequent lower amplitude outbursts for higher mass transfer
rate regardless of model, as illustrated by comparing the two mass transfer
rates shown in Figure~\ref{fig:lightcurves} for each of our three models.
Hence merely reducing the mass transfer rate in the DIMRI model can
increase the outburst amplitude and restore quiescent phases.
If one sets \Mtr as low as $\Mtr=6\times10^{13}$ g/s in the DIMRI (with
$\alpha_{\rm ml}=6$) the elapsed time between two consecutive outbursts starts
to be longer and also the outburst amplitude rises (see
Table \ref{Tab:LC_param} and Figure \ref{fig:DIMRI-6e13}).  Note that for
such a low \Mtr, the discs in DIM and DIMa become cold and stable.

There is yet another alternative way to restore
quiescence in the DIMRI light curves.  Observed X-ray fluxes in
quiescent dwarf novae are far too large compared to what models predict.
A solution to this problem is a truncation of the inner disc.
Numerical calculations by \citet{H00} and \citet{SHL} confirm that truncating
the inner
disc has substantial influence on the dwarf novae lightcurves and may solve
the discrepancy between observations and theory.
This truncation may be caused by the magnetic field of the primary white dwarf.
A white dwarf magnetic field in the range $10^4-10^7$G (which translates to
a magnetic moment $\mu\approx1-10^3 \times10^{30}$~G~cm$^3$)
is sufficient for the magnetic pressure close to the white dwarf to exceed
the gas and ram pressures of the infalling matter during the quiescent phase
of the dwarf nova cycle. The inner disc radius is therefore pushed away from
the white dwarf to a radius $R_M$ (e.g. \citealt{FKR}):
\begin{equation}
R_{\mathrm{in}}=R_{\mathrm{M}}=9.8\times10^8\Mtr_{15}^{-2/7}M_1^{-1/7}\mu_{30}^{4/7} \mathrm{cm}
\label{eq:RM}
\end{equation}
where $\mu_{30}$ is the magnetic moment in units of
$10^{30}\,\,{\rm G\, cm}^{3}$, $M_{1}$ is the mass of the primary in
solar masses and $\dot{M}_{15}$ is the mass accretion rate in units of
$10^{15}\,\,\mathrm{g/s}$.
During outburst the situation changes, as the higher mass accretion rate
sharply increases the ram pressure of matter which then dominates the
magnetic pressure, and the inner edge of the disc approaches the surface
of the white dwarf.
Taking into account the variation in inner disc radius according to
Eq.(\ref{eq:RM}) restores the quiescent
phase in the simulated DIMRI light curves
(compare Figures~\ref{fig:lightcurves}f and \ref{fig:amltrunc}).

\section{Discussion}

There are four main aspects of observed dwarf nova light curves which need to
be reproduced in order to have a successful theoretical model: quiescence
duration, outburst amplitude, outburst duration, and shape.
For the most part our MRI based models DIMa (which differ from the DIM models only by incorporating the
MRI simulation based $\alpha\left(\Teff\right)$) and DIMRI can reproduce these attributes, with the notable exception of reflares. These reflares are the result of two contributing factors.
Namely, the small ratio of the surface densities at
the ends of the lower and upper branches, \QSc, and the low value of $\alpha$ found high up on the upper branch. This low $\alpha$ causes an excess of surface mass density $\Sigma$ in the inner disc. Consequently, as a cooling front propagates inward through the disc it accumulates substantial mass in the post-front region, and due to the small \QSc, this post-front excess in $\Sigma$ easily surpasses the critical value, \Scm, and initiates a reflare. The reflares created by this series of events tend to prolong the decay from outburst. We have found two mechanisms which help to hasten this elongated decay from outburst: reducing the mass transfer rate and truncation of the inner disc by a white dwarf magnetic field
(Figures \ref{fig:DIMRI-6e13}-\ref{fig:amltrunc}). However, these tweaks are merely attempts to treat the symptoms caused by the greater underlying issues mentioned above\footnote{Notice, however, that disc truncations is independently required by dwarf nova observations\citep[see e.g.,][]{L01}.}.

It is important to remember that \QSc is determined by the physics
of both the upper and lower branches, and the lower branch physics is very
uncertain.
The simulations were produced with the ideal MHD
and flux limited diffusion approximations. While these
are reasonable on the upper branch, the lower branch is a different story.
Optical depths along the lower branch are low ($\tau\lesssim5$) bringing into question our usage of the flux limited diffusion approximation.
Moreover,
non-ideal effects, particularly resistivity
and Hall effects \citep[][Appendix \ref{sec:non-ideal}]{BAI14,LES14} are important on
the lower branch where the ionization fraction is low.
This strongly motivates the need to pursue non-ideal MHD simulations
of the lower branch, and
as in protostellar
discs \citep{IGE99}, it may be necessary to account for
irradiation by ionizing x-rays from the boundary layer between the white dwarf
and the accretion disc, because any source of ionization has the potential to significantly modify non-ideal MHD effects.
MRI turbulence may only exist in the irradiated
surface layers, possibly leaving a magnetically driven laminar flow in
the resistive ``dead zone" interior \citep{TUR08}.
Small amounts of hydrodynamic angular momentum transport associated with
thermal convection \citep{LES10} may also exist.
However, it may turn out that none of these local mechanisms is sufficient to
explain the quiescent state, and that something involving more global physics
is required. For example, recent isothermal and adiabatic global simulations
by \citet{JU16} suggest that spiral waves excited by the tidal field of the
donor star may contribute significantly to the angular momentum transport
in the quiescent state, and this physics cannot be captured by a local
stress-pressure relation.
All this speculation, again highlights the uncertainty in \QSc, and the need for the inclusion of non-ideal effects on the lower branch.

Even with this uncertainty, however, we have found that
enhancement of $\alpha$ only near the end of the upper branch, where thermal convection happens,
is capable of producing outbursts. This outcome is non-trivial and suggests that
convection plays a major role in the outbursts of dwarf novae, and
while important, convection may not be the end of the story. The low values of $\alpha$ high up on the upper branch as found by \citet{HIR14}, are a contributing factor in the appearance of reflares.
As mentioned early on in this paper, MRI simulations with net vertical
magnetic flux have larger $\alpha$ values \citep{HAW95,SAN04,PES07}.
While we have considered a way of increasing $\alpha$ with zero net vertical
magnetic flux by the effects of hydrodynamic convection on MRI turbulence,
there is no reason to view these explanations as mutually exclusive.
Accordingly, simulations examining how net vertical magnetic flux modifies
convection and the associated enhancement of $\alpha$ found in \cite{HIR14}
would be useful and may help alleviate reflares.

It is also important to note that we have not successfully incorporated all aspects
of the physics observed in the simulations in our attempts to model outburst
light curves here.
The MRI-based physics that we have succeeded in
incorporating into the DIM has not produced S-curves that completely agree
with the simulation data (cf. Figure~\ref{fig:s-curve}).
Both the upper and lower branches of the MRI simulation S-curve are extended slightly further in $\Sigma$ compared to DIMRI. This implies that DIMRI should have a larger \QSc which in turn would increase the amplitude and the
quiescence duration and may reduce or even alleviate the reflares.

Moreover, it is clear that the MRI simulations exhibit some time dependent
behavior (e.g. intermittent convection) that the DIM simply is incapable of
handling.
During the convective epochs the \citet{SS73} $\alpha$ parameter is enhanced and also $\alpha_{\rm ml}\sim 1.5$. To re-emphasize this, $\alpha$ is significantly higher and $\alpha_{\rm ml}$ significantly lower during the intermittent convective epochs than their respective time averaged quantities.
This has a significant impact on the evolution of an annulus as it transitions
from quiescence to outburst\footnote{
We note that our shearing box simulations are local not only in the radial direction, but also in the azimuthal direction. Therefore, the time dependent behavior may be less manifest in observations where azimuthal variations are averaged.}.
Thermally unstable MRI simulations that are
heating towards the hot branch are fully convective, with no intermittency.
Therefore, when an annulus first makes it to the hot branch it will be fully
convective and ``see" an instantaneous S-curve which is characterized by the $\alpha$ and $\alpha_{\rm ml}$ seen during convective epochs.
This annulus will then evolve and after a few thermal times become
better characterized by the time averaged values for $\alpha$ and
$\alpha_{\rm ml}$ and the associated ``long-term" S-curve.  Global MRI
simulations of the propagation of heating and cooling fronts would obviously
be of great interest, but are not yet viable for the dwarf nova problem.

We finally note that the quiescent state is also a problem for the DIM models (i.e. DIM, DIMa,
DIMRI) presented here and historically in the standard DIM; see e.g.
\citet{L01}. Our lower branch DIM models with $T_{\rm eff}<3000$~K have low optical depths\footnote{
This does not occur in the classical DIM, because without the MRI-based
constraints we have imposed in this paper \ac and \ah can be tuned to avoid
this problem.}
($\tau\lesssim5$) which leads to inaccuracies, most notably $\sim10-50\%$ of
mass lying outside of the vertical boundary condition. However, until non-ideal
effects are included in the lower branch simulations, it is futile to try
and bring the simulation data and DIM models into better agreement here.
This impacts the details of outburst lightcurves, most notably
the occurrence of reflares. While these reflares are not observed in the
outbursts of standard dwarf novae, we nevertheless recover outburst time scales
and amplitudes comparable to those observed, implying that the
simulation-based models of the upper branch are in good agreement with observations.

\section{Conclusions}

We have incorporated the convection-induced enhancement of $\alpha$
close to the hydrogen ionization temperature that was discovered in MRI
simulations by \citet{HIR14} into the DIM.
This, for the first time, places the early inference by \citet{MO,SMAK84,MMH84} that $\alpha$ had to be
larger in outburst in dwarf novae on a strong and clear theoretical foundation
based on MRI turbulence.  We have also shown how to incorporate aspects
of the time-averaged vertical structure (dissipation profiles, intermittent
convective heat transport) into the DIM.  With suitable parameter choices
and/or truncation of the inner disc by a white dwarf magnetic field, the
resulting lightcurves are able to produce outburst and quiescent durations,
as well as outburst amplitudes, that are consistent with observations of
dwarf novae.  Further work to actually fit particular dwarf nova systems
with this model might be worthwhile.

A generic feature of our MRI-based $\alpha$-prescription is the
appearance of reflares.
These reflares are caused by the confluence of low $\alpha$'s high up on the upper branch and the small ratio of critical surface mass densities \QSc. Our biggest uncertainty in this work is the physics of the lower branch and consequently \QSc. This uncertainty primarily stems from the exclusion of important non-ideal MHD effects associated with the largely
electrically neutral plasma on the lower branch.  This gives us a clear motivation and direction for future work to address these inaccuracies in new simulations and to obtain a more realistic understanding of the lower branch and its end.

Additionally, while
we have tried to incorporate MRI-based physics into the DIM, this has only
been done for annuli in thermal equilibrium.  We still do not understand
how MRI physics might affect the propagation of heating and cooling fronts,
and it is these that are responsible for reflares.
MRI simulations in a more global geometry that can track the propagation of such fronts would of course be very illuminating, though such simulations appear to be challenging with current resources.

Finally, it is important to remember that reflares are also a common
problem even in the standard DIM, where one is free to choose values of
\ah and \ac to try and get rid of them (e.g. \citealt{L01}).  This freedom
is of course an illusion, as the stresses are actually determined by
the underlying physics of turbulence in the disc.  The fact that applications
of our MRI simulation results to the DIM give rise to reflares helps sharpen
this problem by relating it more to the fundamental physics responsible
for stresses in the disc.

\section*{Acknowledgments}

We thank the anonymous referee for a constructive report that led to
significant improvements in this paper.
We also wish to acknowledge the seminal role of Jean-Marie Hameury and
Guillaume Dubus in developing the numerical methods necessary to accurately
model lightcurves within the DIM.

Additionally, we also thank Jean-Marie Hameury for constructive feedback on this work.
This research was supported by the United States NSF
grant AST-1412417, the Polish NCN grants UMO-2013/08/A/ST9/00795 and UMO-2015/19/B/ST9/01099, and by Japan JSPS KAKENHI Grants 24540244 and
26400224, joint research project of ILE, Osaka University.
JPL was supported in part by a grant from the
French Space Agency CNES.  OB thanks the International Space Science Institute
in Bern for its support while this paper was being completed.
Numerical calculations were partly carried out on the Center for Scientific
Computing from the CNSI, MRL: an NSF MRSEC (DMR-1121053) and NSF CNS-0960316,
and the Cray XC30 at CfCA, National Astronomical Observatory of Japan.

\bibliography{citations}

\appendix

\section{Non-ideal MHD Effects in Dwarf Nova Quiescence}
\label{sec:non-ideal}
Here we examine non-ideal MHD effects due to thermal ionization only (i.e. neglecting irradiation) and their possible importance in dwarf nova quiescence. By non-ideal MHD effects, we are referring to Ohmic dissipation (i.e. resistivity), the Hall term, and ambipolar diffusion.
The full induction equation including these terms can
be written as \citep{LES14}\footnote{
Here we use Gaussian units, thus our presentation differers by factors of $\sqrt{4\pi}$ compared to that of \citet{LES14}.
}
\begin{equation}
\dfrac{\partial{\bf B}}{\partial t}
-{\bnabla}\times\left({\bf v}\!\times\!{\bf B}\right)
\!=\!-\dfrac{\bnabla}{4\pi}
\times\left(
\eta_{\rm O}{\bf J}
+\eta_{\rm H}{\bf J}\!\times\!{\bf e_b}
-\eta_{\rm A}{\bf J}\!\times\!{\bf e_b}\!\times\!{\bf e_b}
\right),
\end{equation}
where ${\bf J}=4\pi\bnabla\times{\bf B}$ and ${\bf e_b}={\bf B}/\left|{\bf B}\right|$.
The terms on the right hand side are respectively
the Ohmic resistive term (O), the Hall term (H), and the ambipolar diffusion term (A), where the $\eta$ coefficients are the respective diffusivity coefficients. The Hall term is caused by a velocity difference between electrons (e) and heavy cations (i), while Ohmic dissipation and ambipolar diffusion are caused by collisions between neutrals (n) and electrons or cations respectively.
Thus, it is worthwhile briefly discussing
the relevant quantities involving these collisions.
The momentum rate coefficients for ion-neutral
and electron-neutral collisions are
$<\sigma v>_{\rm ni}=1.9\times10^{-9}$~cm$^3$~s$^{-1}$ \citep{DRA11} and
$<\sigma v>_{\rm ne}=8.3\times10^{-10}T^{1/2}$~cm$^3$~s$^{-1}$ \citep{DRA83}, respectively. The ion-neutral drag coefficient is therefore
\begin{equation}
\gamma=\dfrac{<\sigma v>_{\rm ni}}{m_{\rm n}+m_{\rm i}}
=2.7\!\times\! 10^{13}
\left(\dfrac{m_{\rm n}+m_{\rm i}}{41.37\,m_{\rm u}}\right)^{-1}
\!{\rm cm}^3\,
{\rm s}^{-1}\,{\rm g}^{-1},
\end{equation}
where we have taken $m_{\rm n}=2.37\,m_{\rm u}$ (the mean molecular weight for our
abundances, taking all the hydrogen to be molecular), and $m_i=39 \,m_{\rm u}$
\citep[K$^+$, appropriate for thermal ionization;][]{BAL01}\footnote{
Note that \citet{BLA94} adopted $m_{\rm i}=30m_{\rm u}$, appropriate for the interstellar medium; while \citet{SAN00} adopted $m_i=24 \,m_{\rm u}$ (Mg$^+$) appropriate for ionization in accretion discs due to irradiation.}.
The diffusivity coefficients for Ohmic, Hall, and ambipolar diffusion are as follows \citep[in the absence of dust;][]{BAL01,WAR07,LES14}:
\begin{align}
\eta_{\rm O}&=\dfrac{c^2m_{\rm e}}{4\pi e^2} \dfrac{n_{\rm n}}{n_{\rm e}} \mean{\sigma v}_{\rm ne}\\
\eta_{\rm H}&=\dfrac{Bc}{4\pi en_{\rm e}}=\sqrt{\dfrac{\rho}{4\pi}}\dfrac{v_{\rm A}c}{n_{\rm e}e}\\
\eta_{\rm A}&=\dfrac{B^2}{4\pi\gamma_{\rm i}\rho\rho_{\rm i}}
=\dfrac{v_{\rm A}^2}{\gamma_{\rm i}\rho_{\rm i}},
\end{align}
where $v_{\rm A}=B/\sqrt{4\pi\rho}$ is the Alfv\'{e}n speed.

By introducing
the isothermal sound speed $c_{\rm s}=\sqrt{kT/\mu m_{\rm u}}$ we can examine the ratios
\begin{align}
\dfrac{\eta_{\rm H}}{\eta_{\rm O}}
&=\sqrt{4\pi}
\dfrac{v_{\rm A}}{c_{\rm s}}
\sqrt{\dfrac{n_{\rm tot}}{n_{\rm n}}}
\dfrac{\sqrt{kT}}{\mean{\sigma v}_{\rm ne}}
\dfrac{e}{cm_{\rm e}} n_{\rm n}^{-1/2}\\
\nonumber
&=1.14\dfrac{v_{\rm A}}{c_{\rm s}}
\sqrt{\dfrac{n_{\rm tot}}{n_{\rm n}}}
\left(\dfrac{m_{\rm u} n_{\rm n}}{10^{-6}\text{ g cm}^{-3}}\right)^{-1/2}
\\
\dfrac{\eta_{\rm A}}{\eta_{\rm H}}
&=\sqrt{4\pi} \dfrac{n_{\rm e}}{n_{\rm i}} \dfrac{v_{\rm A}}{c_{\rm s}} \dfrac{e}{\gamma_{\rm i}cm_{\rm i}\sqrt{\rho}} \sqrt{\dfrac{kT}{\mu m_{\rm u}}}\\
\nonumber
&=6.61\!\times\! 10^{-3}
\dfrac{n_{\rm e}}{n_{\rm i}}
\dfrac{v_{\rm A}}{c_{\rm s}}
f_{\rm M}
\left(\rho_{-6}\right)^{-1/2}
\left(\dfrac{T}{3000\text{ K}}\right)^{1/2}
\\
f_{\rm M}&\equiv
\left(\dfrac{m_{\rm i}}{39m_{\rm u}}\right)^{-1}
\left(\dfrac{m_{\rm n}\!+\!m_{\rm i}}{41.37\,m_{\rm u}}\right)
\left(\dfrac{\mu}{2.37}\right)^{-1/2},
\end{align}
where $\rho_{-6} = \rho / 10^{-6}$ g cm$^{-3}$, and $f_{\rm M}=1$ for the abundances and masses we assume here.
From this we immediately
conclude that the Hall term will likely be important whenever the Ohmic dissipation is important, as we are typically talking about densities and temperatures in this range. However, ambipolar diffusion is negligible compared to both of the other two non-ideal effects.

\begin{figure}
	\centering
	\includegraphics[width=\linewidth]{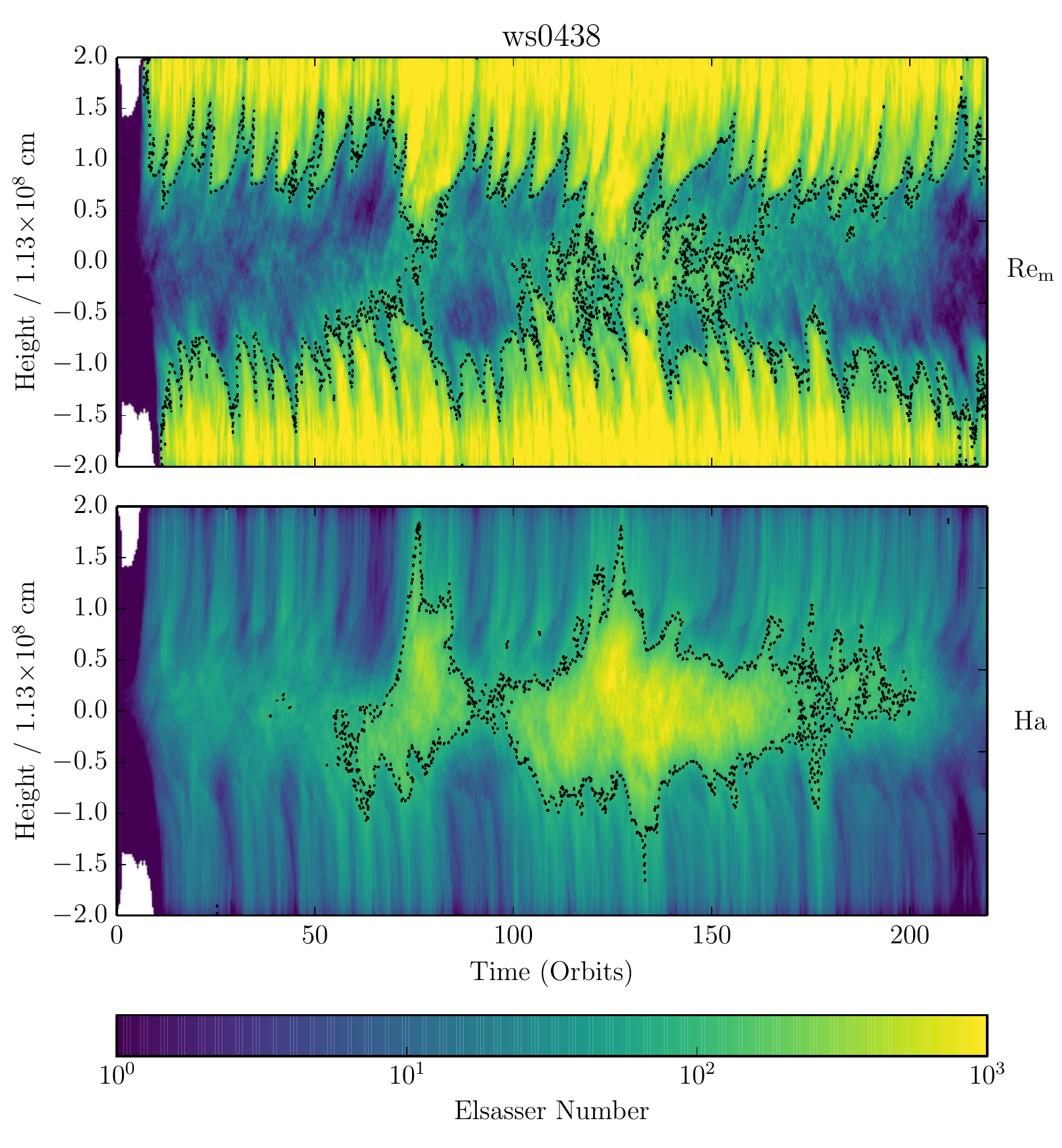}
	\caption{
	Horizontally averaged Elsasser numbers for the hottest lower branch simulation (ws0438). Lower Elsasser number corresponds to non-ideal MHD effects being more important. Re$_{\rm m}$ (Eq. \ref{eqn:Rm}, top panel) corresponds to Ohmic dissipation and Ha (Eq. \ref{eqn:Ha}, bottom panel) corresponds to the Hall term. Previous works \citep[e.g.][]{HAW96,LES07} suggest that Ohmic dissipation is important for Re$_{\rm m} \lesssim 100$; we have therefore plotted a dotted black contour at Re$_{\rm m}$, Ha $=100$. Non-ideal MHD effects are thus likely important for the majority of this simulation.
	}
	\label{fig:nonideal}
\end{figure}

To determine how important the Ohmic and Hall terms are, it is useful to examine the following dimensionless Elsasser numbers which can be computed from simulation profiles:
\begin{align}
\label{eqn:Rm}
{\rm Re}_{\rm m}&\equiv\dfrac{v_{\rm A}^2}{\Omega \eta_{\rm O}}\\
\label{eqn:Ha}
{\rm Ha}&\equiv\dfrac{v_{\rm A}^2}{\Omega \eta_{\rm H}}
\end{align}
For the hottest lower branch simulation (ws0438) we have computed the horizontally averaged Elsasser numbers (see Figure \ref{fig:nonideal}).
Previous works \citep[e.g.][]{HAW96,LES07} show that
Ohmic dissipation is important when Re$_{\rm m}$ is less than a few hundred and much of ws0438 has Elsasser numbers below 100 (the dotted black contour of Figure \ref{fig:nonideal}). The other lower branch simulations are colder than ws0438 and are therefore more susceptible to non-ideal effects, implying that non-ideal MHD effects are important in dwarf nova quiescence. Future work should therefore include Ohmic dissipation and the Hall term in our
simulations, but the ambipolar diffusion term is probably not necessary.

\section{Simulation Parameters}
The parameters for 3D MHD shearing-box simulations which appear in this paper but not in \citet{HIR14} are listed below in Table \ref{table:sims}.
\begin{table*}
\footnotesize
\setlength{\tabcolsep}{4pt}
\begin{tabular}{cccccccccccccccccc}
\hline
\hline \\[-2.3ex]
$T_{\rm eff}$ & $\Sigma$ & $\alpha$ & $\tau_{\rm tot}$ & $T_{\rm eff,0}$ & $\Sigma_0$ & $\alpha_0$ & $\beta_0$ & $h_0/10^8$ & $N_x$ & $N_y$ & $N_z$ & $L_x/h_0$ & $L_y/h_0$ & $L_z/h_0$ & $L_z/h_P$ & $t_{\rm th}$ & $t_{\rm max}$\\[.2ex]
\hline
\hline \\[-2.3ex]
\multicolumn{18}{c}{$R=1.25 \! \times \! 10^{9}$ cm}\\[.2ex]
\hline
8932 & 10.7 & 0.090 & 5876 & 8709 & 10.9 & 0.042 & 10 & 0.226 & 32 & 64 & 384 & 0.50 & 2.00 & 6.00 & 15.2 & 13.8 & 189\\
9427 & 12.2 & 0.068 & 4910 & 7943 & 12.5 & 0.026 & 10 & 0.210 & 32 & 64 & 384 & 0.50 & 2.00 & 6.00 & 11.5 & 17.1 & 114\\
9698 & 14.7 & 0.055 & 5162 & 7943 & 14.9 & 0.020 & 10 & 0.221 & 32 & 64 & 384 & 0.50 & 2.00 & 6.00 & 10.9 & 19.7 & 118\\
17405 & 85.8 & 0.037 & 6985 & 16218 & 86.4 & 0.026 & 100 & 0.447 & 32 & 64 & 512 & 0.50 & 2.00 & 6.00 & 13.4 & 24.4 & 133\\
19471 & 106 & 0.042 & 6798 & 16218 & 107 & 0.020 & 10 & 0.462 & 32 & 64 & 512 & 0.50 & 2.00 & 6.00 & 13.0 & 20.9 & 139\\
\hline \\[-2.3ex]
\multicolumn{18}{c}{$R=4.13 \! \times \! 10^{9}$ cm}\\[.2ex]
\hline
7546 & 36.7 & 0.134 & 17221 & 8709 & 37.2 & 0.067 & 10 & 1.55 & 32 & 64 & 384 & 0.50 & 2.00 & 6.00 & 22.7 & 11.0 & 165\\
8079 & 40.4 & 0.099 & 16800 & 9549 & 40.6 & 0.081 & 10 & 1.71 & 32 & 64 & 384 & 0.50 & 2.00 & 6.00 & 19.9 & 12.8 & 114\\
8542 & 45.2 & 0.076 & 14139 & 9549 & 45.5 & 0.070 & 10 & 1.75 & 32 & 64 & 384 & 0.50 & 2.00 & 6.00 & 16.4 & 15.1 & 122\\
9324 & 52.0 & 0.069 & 9283 & 9549 & 52.5 & 0.058 & 10 & 1.80 & 32 & 64 & 384 & 0.50 & 2.00 & 6.00 & 13.9 & 15.9 & 123\\
\hline
\end{tabular}
\caption{Parameters for simulation runs.
$T_{\rm eff}$, $\Sigma$, $\alpha$, and $\tau_{\rm tot}$ are the time averaged effective temperature in K, column mass density in g cm$^{-2}$, \citet{SS73}
$\alpha$-parameter, and optical depth respectively.
$T_{\rm eff,0}$, $\Sigma_0$, $\alpha_0$, and $\beta_0$ are the initial condition values where $\beta_0$ is the initial ratio of gas to magnetic pressure.
$h_0/10^8$ is the simulation length unit over $10^8$ cm.
$N_x$, $N_y$, $N_z$ are the number of grid cells in the $x$, $y$, and $z$ directions respectively.
$L_x/h_0$, $L_y/h_0$, $L_z/h_0$ are dimensions of the simulation domain in the $x$, $y$, and $z$ directions respectively.
$L_z/h_P$ is the height of the simulation domain over the time averaged pressure scale height.
$t_{\rm th}$ is the time averaged thermal time. $t_{\rm max}$ is the maximum runtime used for time averaging.
}
\label{table:sims}
\end{table*}

\label{lastpage}

\end{document}